# Quantum state transfer between valley and photon qubits

Ming-Jay Yang[1], Han-Ying Peng[2], Neil Na[3], and Yu-Shu Wu[1,2]

[1] Institute of Electronics Engineering, National Tsing-Hua University, Hsinchu 30013, Taiwan
[2] Department of Physics, National Tsing-Hua University, Hsinchu 30013, Taiwan
[3] Artilux Inc., Zhubei City, Hsinchu 30288, Taiwan

The electron-photon interaction in 2D materials obeys the rule of "electron valley – photon polarization" correspondence. At the quantum level, such correspondence can be utilized to entangle valleys and polarizations and attain the transfer of quantum states (or information) between valley and photon qubits. Our work presents a theoretical study of the interaction between the two types of qubits and the resultant quantum state transfer. A generic setup is introduced, which involves optical cavities enhancing the electron-photon interaction as well as facilitating both the entanglement and un-entanglement between valleys and polarizations required by the transfer. The quantum system considered consists of electrons, optically excited trions, and cavity photons, with photons moving in and out of the system. A wave equation based analysis is performed, and analytical expressions are derived for the two important figures of merits that characterize the transfer, namely, yield and fidelity, allowing for the investigation of their dependences on various qubit and cavity parameters. A numerical study of the yield and fidelity has also been carried out. Overall, this work shows promising characteristics in the valley-photon state transfer, with the conclusion that the valley-polarization correspondence can be exploited to achieve the transfer with good yield and high fidelity.

## I. INTRODUCTION

The valley degree of freedom in electrons has recently attracted a lot of attention, in particular in 2D hexagonal materials such as graphene [1-3] and transition metal dichalcogenides (TMDCs) [4-7]. In addition to its unique electromagnetic properties [1-3], this degree of freedom also manifests novel optical behaviors [4-8]. Altogether, a wide spectrum of exciting opportunities are created for the valley based electronics known as valleytronics.

In the class of 2D hexagonal materials [9-12], the three-fold rotational symmetry comprises the physical root of the intriguing valley-dependent physics. Basically, in the presence of an energy gap, the constraint of symmetry on electron states results in intrinsic, unit-cell-scale orbital angular momenta with opposite signs for electrons in the Dirac valley doublet at K and K′ of the Brillouin zone [2, 8]. In close analogy to the ordinary electron spin, where opposite angular momenta characterize up and down states, a valley pseudospin thus emerges with index given by, for example, $\tau_v = +1$ for K and -1 for K′, which fully qualifies for the role of a quantum bit carrier as the electron spin does. In particular, concrete proposals have been given of a valley-based approach to the fundamental unit, namely, a qubit, for the application of quantum information processing [13-17] based on quantum dots (QDs) [18-20]. In the case of graphene, for example, the valley degree of freedom can be incorporated to expand an electron spin qubit to a spin-valley qubit [14] or, one can freeze out the spin and construct a valley-pair qubit out of a pair of quantum dots, with each QD localizing an electron and subject to the modulation of both an external magnetic field and electrical gates for qubit manipulation [13]. Demonstration has been given in the latter case showing the satisfaction of DiVincenzo criteria for universal quantum computing.

In a way similar to semiconductor electron spin qubits [21], valley qubits can interact with photon qubits with a good coupling strength. Such an inter-qubit interaction is characterized by several promising features. First, the 2D materials of interest have direct band gaps at Dirac points allowing for strong, vertical optical transitions. In the case of graphene, a large optical matrix element $\sim ev_F A$ exists for the transition, due to the sizable $v_F$ ($v_F$ = Fermi velocity $\sim 10^6$ m/sec, $e$ = electron charge, $A$ = vector potential). In addition, valley qubits can be integrated with cavities or waveguides of planar photonic structures [22,23] for an enhancement of the electron-photon (e-ph) interaction. In fact, control of the interaction in 2D materials using cavities has been demonstrated under both strong [24] and weak [25] coupling regimes, which paves the path for implementing photon-valley interfaces required for valley-involved optoelectronics.

The valley-photon interaction physics is significantly enriched by the existence of a valley-dependent selection rule for optical transitions. Due to the presence of finite valley and photon angular momenta, the law of angular momentum conservation leads to interband transitions that are governed by the valley-dependent selection rule as shown in **Figure 1** with the involved photons being circularly-polarized. Because of this selection rule, an approximate one-to-one correspondence exists between circular polarizations of pumping (emitted) photons and valley states of excited (recombining) electron-hole pairs [4-8]. Experimentally, the past few years have seen great strides in the field of optovalleytronics, in spin-valley pumping by optical excitations that utilize the selection rule [4-7]. From the perspective of quantum information processing, the valley-polarization correspondence is also of great interest. Since this correspondence implies the existence of a natural quantum state transfer (QST) between photon and valley qubits [26], a theoretical and experimental raise of its utilization via such QST to the quantum information processing level of applications would fulfill the full potential of the correspondence.

The QST is a coherent quantum process where, via the e-ph interaction, the photon state information can be extracted and stored in a valley qubit, and the reverse process that moves the state information of a valley qubit back to a photon can as well occur. Such valley-photon QST provides an analogy to the well-studied spin-photon QST [27]. In the latter case, for example, a coherent single photon detection is realized via the process of photon absorption, electron-hole pair generation, and hole extraction, leaving the polarization state of the photon totally encoded in the spin of the excited electron. On the other hand, the valley-photon QST also constitutes an interesting contrast to the spin-photon QST, with the following fundamental difference existing between the two.

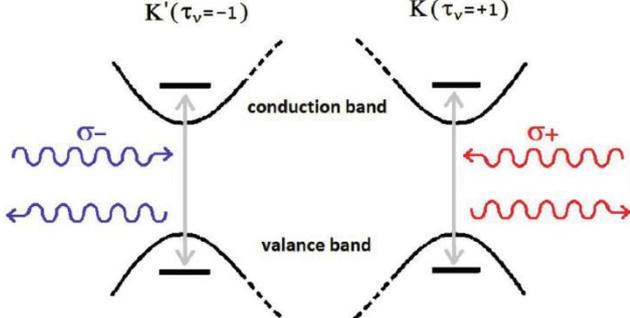

**Figure 1** Approximate selection rule for the interband optical transition in gapped graphene. $\sigma_{+(-)} = \sigma_x +(-) i\sigma_y$ denotes circular polarization states in the graphene plane.

In the electron spin case, a single spin forms the simplest qubit in the class of spin qubits and hence has been the focus of spin-photon QST studies; whereas, in the valley case, a psuedospin qubit must consist of two valley pseudospins with the states being given by the so-called "decoherence-free singlet / triplet states" [28]. Therefore, this valley-pair qubit makes a natural choice for valley-photon QST, and the corresponding study would, from the scientific perspective, constitute an attempt at understanding the nature of QST between photons and the class of "decoherence-free" qubits.

From the application point of view, the study of valley-photon QST would also advance the field of quantum technology in 2D materials. For example, this specific QST would facilitate the development of valley-involved quantum tomography, where one can reconstruct the valley (photon) state via a measurement done on the photon (valley) state, when the latter is easier to characterize than the former. In addition, it is well suited to the application of quantum communications (QCs) [26], with a potential to realize the quantum repeater (QR) protocol as follows. A QR extends the distance of QCs beyond that of photon attenuation, by an iterative process involving the QST from photons to quantum memories (static forms of qubits), and eventually sets up a global entanglement required for quantum teleportation [29,30]. For such an application, the QST involved is required to be as faithful as possible, since any state distortion during the transfer would lead to a corresponding reduction in the degree of entanglement and hence loss of fidelity in the information teleported. From such a perspective, quantum memories can be implemented with only a few physical systems. For instance, coherent state transfer and control in atomic and ionic systems has made a great progress [31-34]; superconducting qubits with transmission line cavities in circuit quantum electrodynamic setup offer the possibility of large-scale, fault-tolerant quantum information processing with integrated qubits [35]; the electronic spin triplet ground state in a nitrogen-vacancy defect exhibits a promising long decoherence time [36,37]; for quantum dot-confined electron spin qubits, the state transfer from photon polarization to electron spin using optically-active semiconductor QDs has been extensively studied and demonstrated [38,39]. Like these systems, valley qubits also carry properties that qualify them for quantum memories. For valley qubits, one represents logical 0 and 1 with the low energy sector of states where the dynamical variables consist only of valley pseudospins of electrons. Such qubits have the advantage that the large wave vector difference between K and K' valleys stabilizes the qubit state and provides a good coherence protection, with typical valley relaxation time for such qubits to be in the range of $10^{-6} \sim 10^{-3}$ sec [13,26]. However, it should be noted that the simplest realization of a valley qubit with a single valley pseudospin may be quite challenging. With the qubit state space {0, 1} represented by the two-state space {K, K'} of the pseudospin, such a qubit faces the issue of being difficult to transform between, for example, K and K' via controllable means due to the wave vector difference between them. This difficulty would have to be overcome in applications involving qubit state manipulations as in the case of QRs when setting up the global entanglement [29,30]. On the other hand, a valley-pair qubit with two valley pseudospins resolves such an issue[13]. In this case, spin and orbital degrees of freedom are removed by a magnetic field-induced spin quantization and the QD confinement-induced localization, respectively. The remaining degrees of freedom, i.e., the two localized pseudospins, interact with each other via a spin exchange type coupling and form the two maximally entangled states, namely, the valley singlet state $|Z_S\rangle = \frac{1}{\sqrt{2}}(|K_L K'_R\rangle - |K'_L K_R\rangle)$ (with subscripts L and R denoting the left and right QDs, respectively) and the valley triplet state $|Z_{T_0}\rangle = \frac{1}{\sqrt{2}}(|K_L K'_R\rangle + |K'_L K_R\rangle)$. The two states can represent 0 and 1, respectively, and a single-qubit transformation in the {0, 1} space can be performed without any valley flipping [13] as described in the following in terms of the Bloch sphere representation of qubit states. First, the exchange coupling between the QDs can be electrically controlled to rotate the qubit around one axis of the sphere. Second, a mechanism called valley-orbit interaction (VOI) exists between the valley pseudospin and an in-plane electric field. This field can be induced by electrical gates near the QDs to allow, via the VOI mechanism, for rotation of the qubit around another axis of the sphere. The two forgoing independent rotations can be combined to achieve an arbitrary single qubit manipulation on the time scale of $10^{-9}$ sec [13,26]. Such an electrical gated valley qubit features scalability similar to typical solid state qubits, and can be advantageous to the implementation of quantum error correction (QEC) coding [40,41]. Due to the fact that the QEC represents a single logical qubit with a cluster of physical qubits, construction of the corresponding circuit can be facilitated by the scalability of valley qubits.

The photon-valley QST is a complicated quantum-mechanical problem. It involves a system of valley qubit electrons, optically excited electrons and holes, and photons, with the e-ph coupling existing between the particles. Moreover, the system is open and communicates with the external world via the photonic signal moving in and out of the system. The present work provides an initial, yet semi-quantitative understanding of this complicated problem through an analytical approach based on a set of approximations. It introduces a generic setup for the photon-valley QST that can be optimized for the yield and fidelity, and investigates the QST in a sophisticated quantum-mechanical model beyond what is merely based on the approximate valley-polarization correspondence. Specifically, the setup consists of the valley qubit being placed inside both a distributed Bragg reflector (DBR) based cavity and a photonic crystal (PC) cavity. The DBR cavity serves to enhance the e-ph interaction for the absorption of incoming signal photon by the valley qubit and so facilitates the valley-polarization entanglement. The PC cavity serves to enhance the photon emission from the photo-excited valley qubit as well as project the linear polarization state of the emitted photon and so facilitates the valley-polarization un-entanglement. Through the entanglement and un-entanglement processes, the quantum state information of the incoming photon is shared with and transferred to the valley qubit. A quantum-mechanical analysis is performed for such processes in terms of realistic optical matrix elements and a reasonable phenomenological modeling of damping for both the electron states and cavity modes, yielding quantum mechanical equations that govern the time evolution of various probability amplitudes in the system. These amplitudes are analytically solved to determine the expressions of yield and fidelity - the fidelity measures the fraction of faithful QST per transfer, and the yield describes the fraction of photon-to-quantum memory conversion per incoming signal photon. These two figures are the upmost important parameters that determine the efficiency and resources involved in the QST. Their explicit dependences on various cavity and qubit parameters are derived to facilitate our investigation of the optimal conditions for yield and fidelity. A numerical study of the yield and fidelity is also carried out. Overall, the study shows promising characteristics in the photon-valley QST, with the conclusion that the valley-polarization correspondence can indeed be exploited to achieve such a QST with good yield and high fidelity.

The paper is organized as follows. In **Sec. II**, we first describe the setup designed to both enhance the e-ph interaction and differentiate the incoming and outgoing photons. We then discuss the optical matrix element and provide a description of the photon-valley QST in the setup. In **Sec. III**, we present a quantum-mechanical description of the QST process, solve the quantum mechanical equations, and derive the analytical expressions of yield and fidelity. In **Sec. IV**, based on the expressions of yield and fidelity, numerical results are obtained, and their implications are discussed for the photon-valley QST in the proposed setup. In **Sec. V**, we summarize our findings. In **Appendix**, we provide the mathematical details involved in solving the various probability amplitudes.

## II. PHOTON-VALLEY QST IN CAVITIES

**Sec. II-1** overviews the QST in a simple setup with a single optical cavity. **Sec. II-2** discusses the proposed setup with two optical cavities. **Sec. II-3** discusses the optical matrix element due to the e-ph interaction between a qubit electron and a cavity photon. In **Sec. II-4**, we describe the photon-valley QST in the two-cavity setup. The discussion of QST here intends to provide a qualitative picture of the process, by giving the initial, intermediate, and final states involved, in preparation for the discussion of a more complete quantum-mechanical treatment in **Sec. III**.

### II-1. QST with One Cavity

The principle underlying the valley-photon QST is the unique, approximate valley-polarization correspondence mentioned earlier, which enables a natural QST between valley and photon qubits. **Figure 2** shows a simple, conceptual setup for the QST, where the photon enters an optical cavity, interacts with the valley qubit placed inside, and leaves the cavity. Illustration of the concept of valley-photon QST is given below within this setup.

The photon to valley QST is featured by

i) initialization of the valley qubit into the singlet state $\frac{1}{\sqrt{2}}(|K_L K'_R> - |K'_L K_R>)$,

ii) the incoming signal photon of energy $\hbar\omega_{ph}$ in a generic state $\alpha|\sigma_+> + \beta|\sigma_->$ of mixed circular polarizations ($\sigma_+$ and $\sigma_-$) that carries the quantum information in $\{\alpha, \beta\}$,

iii) enhancement of the e-ph interaction in the cavity,

iv) gate tuning of energy levels in one of the two QDs (taken to be $QD_L$, the left QD, throughout the work) to achieve the trion-generating or -eliminating resonant transitions $\hbar\omega_{ph} + |K_L> \leftrightarrow |K'_{eh,L} K_L>$ and $\hbar\omega_{ph} + |K'_L> \leftrightarrow |K_{eh,L} K'_L>$ in $QD_L$, where $|K'_{eh,L} K_L>$ is the trion consisting of one K-electron, one K'-electron and one K'-hole, for example,

v) gate control to switch off the tunneling coupling between the QDs and freeze the

inter-QD orbital motion, thus eliminating the electron in $QD_R$ (the right QD) from our consideration of the QST process, except for its entanglement with the electron in $QD_L$,

vi) linear polarization ($\sigma_x$ and $\sigma_y$) state projection measurement of the valley qubit-emitted photon.

In the ideal case where the valley-polarization correspondence is exact, the QST proceeds in the following sequence [26]:

$$|\Phi_0> = \frac{1}{\sqrt{2}}(|K_L K'_R> - |K'_L K_R>)$$
$$\otimes (\alpha|\sigma_+> + \beta|\sigma_->)$$

$$\xrightarrow{\text{photon absorption}} |\Phi_1> = \frac{1}{\sqrt{2}}(\beta|K'_{ex,L} K_L K'_R> - \alpha|K_{ex,L} K'_L K_R>)$$

$$\xleftrightarrow{\text{photon emission/absorption}} |\Phi_2>$$
$$= \beta|K_L K'_R> \otimes |\sigma_-> - \alpha|K'_L K_R> \otimes |\sigma_+>$$

$$\xrightarrow{\text{projection onto } \sigma_x/\sigma_y} |\Phi_{3x}> = \beta|K_L K'_R> - \alpha|K'_L K_R> \text{ (if } \sigma_x \text{ detected)}$$
$$|\Phi_{3y}> = \beta|K_L K'_R> + \alpha|K'_L K_R> \text{ (if } \sigma_y \text{ detected)}$$

Note that in the intermediate state $|\Phi_2>$, the photon and the qubit electrons are entangled. The electrons now share the quantum information carried in the amplitudes $\{\alpha, \beta\}$. The entangled photon eventually leaks out of the cavity and a $\sigma_x/\sigma_y$ projection measurement is performed on its linear polarization state. The projection un-entangles the photonic component, leaving the information solely stored in the valley-pair state ($\Phi_{3x}$ or $\Phi_{3y}$). If desired, the resultant valley state could be further transformed by standard single qubit manipulations into the combination of valley singlet and triplet states, for example,

$$\beta|K_L K'_R> - \alpha|K'_L K_R>$$
$$\to \alpha\frac{1}{\sqrt{2}}(|K_L K'_R> - |K'_L K_R>) + \beta\frac{1}{\sqrt{2}}(|K_L K'_R> + |K'_L K_R>)$$

thus storing $\{\alpha, \beta\}$ in the robust, "decoherence-free" valley state.

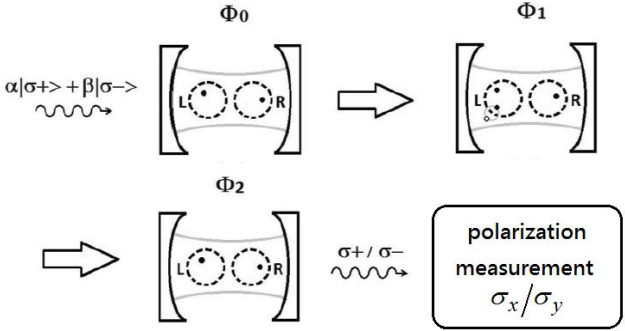

**Figure 2** The QST from a photon qubit to a valley-pair qubit in an optical cavity. Dashed circles – quantum dots; black/hollow dots – electrons/holes.

We note that the existence of bound states of excitons and trions are not required for the valley-photon QST. The terminology of trions is used in our work just to indicate the presence in the QST of intermediate states consisting of two electrons and one hole with Coulomb interaction among them.

The reverse process of valley to photon QST can be similarly achieved. We replace the initial state by

$$|\Phi_0> = (\beta|K_L K'_R> - \alpha|K'_L K_R>)$$
$$\otimes \frac{1}{\sqrt{2}}(|\sigma_+> + |\sigma_->)$$

with the quantum information $\{\alpha, \beta\}$ now encoded into the valley-pair qubit. The inter-qubit interaction leads next to the following state evolution:

$$\xrightarrow{\text{photon absorption}} |\Phi_1> = \frac{1}{\sqrt{2}}(\beta|K'_{ex,L} K_L K'_R> - \alpha|K_{ex,L} K'_L K_R>)$$

$$\xleftrightarrow{\text{photon emission/absorption}} |\Phi_2>$$
$$= \beta|K_L K'_R> \otimes |\sigma_-> - \alpha|K'_L K_R> \otimes |\sigma_+>$$

Now, instead of measuring the photon, we measure the valley qubit, and project it onto singlet / triplet states:

$$\xrightarrow{\text{projection onto } Z_S/Z_{T0}} |\Phi_{3S}> = \alpha|\sigma_+> + \beta|\sigma_-> \text{ (if } Z_S \text{ detected)}$$
$$|\Phi_{3T_0}> = -\alpha|\sigma_+> + \beta|\sigma_-> \text{ (if } Z_{T_0} \text{ detected)}$$

which completes the transfer by storing $\{\alpha, \beta\}$ in the photon states $\Phi_{3S}$ or $\Phi_{3T_0}$.

From now on, our investigation will focus on the photon to valley QST, the process actually used in a QR.

## II-2. Setup with Two Cavities

An issue with the single cavity setup in **Figure 2** lies in the existence of a finite probability for the incoming photon to enter the cavity, leak out of the cavity and enter the polarization projection measurement, without ever interacting and becoming entangled with the electron. Since the e-ph entanglement is a necessary condition for a successful QST, a detection of the idler photon would create a false event of QST and, hence, reduce the fidelity. It is therefore desirable to resolve the overlap between the idling and the entangled photons. This could be done in either the spatial or the frequency domains. In this work, we consider a configuration involving two optical cavities with orthogonal cavity axes - one gives vertical (z) optical confinement and the other transverse (x-y) optical confinement, as shown in **Figure 3**, with the qubit sitting at the center of both cavities. The configuration effects an enhancement of the e-ph interaction as well as a differentiation between the paths of incoming and outgoing signal photons, as follows. The first cavity is formed of a pair of DBRs that provide the vertical confinement. It couples a vertically incoming signal photon into a DBR cavity mode and excites a trion in QD$_L$. On the other hand, for the transverse confinement, we envision a defect in a 2D PC of square lattice structure, as shown in **Figure** 3, with the qubit placed at the center of the defect. The defect forms a cavity with partially confined transverse electric (TE) modes. These PC cavity modes have electric fields and wave vectors lying in the plane [42-44] and hence are efficiently e-ph coupled to a QD$_L$ electron in the graphene plane. The coupling induces the trion to emit a photon into a PC cavity mode, which eventually leaks out of the cavity in a nearly in-plane direction and can be picked up by a sensor. Once detected, the photon becomes disentangled from the valley qubit and this completes the state transfer.

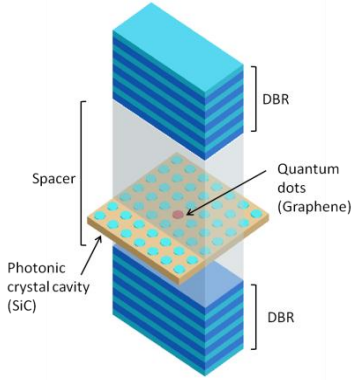

**Figure 3** The proposed hybrid-cavity setup. The valley-pair qubit (red dot in the middle plane) is placed at the center of the PC cavity. It is also vertically confined by the cavity formed with DBRs on both top and bottom sides. The signal photon comes in through the top DBR, excites a trion in QD$_L$, which later radiates a photon moving out of the PC cavity in a nearly horizontal direction into a photon sensor.

### II-3. Optical Matrix Elements

Throughout the work, we take the valley qubit to be embedded in a gapped monolayer graphene (e.g., BN-doped monolayer graphene [45]) such that we can use the simple 2D Dirac theory of monolayer graphene [11,13] in the modeling, in order to simplify the analysis while retaining the essential valley physics. For the QST, we consider the interaction between a QD$_L$ electron and a radiation field of frequency $\omega_{ph}$. The electron in QD$_L$ is governed by the following 2D Dirac equation [13]

$$\left(H_D^{(0)} + H_A\right)\phi_D = i\hbar\partial_t\phi_D,$$

$$H_D^{(0)} = \begin{pmatrix} \Delta(\vec{r}) + V(\vec{r}) & v_F \hat{p}_- \\ v_F \hat{p}_+ & -\Delta(\vec{r}) + V(\vec{r}) \end{pmatrix}, H_A = \begin{pmatrix} 0 & ev_F A_- \\ ev_F A_+ & 0 \end{pmatrix},$$

$$\phi_D = \begin{pmatrix} \varphi_A \\ \varphi_B \end{pmatrix}, \quad (1)$$

$$A_\pm = A_x \pm i\tau_v A_y,$$

$$\hat{p}_\pm = p_x \pm i\tau_v p_y = -i\partial_x \pm \tau_v \partial_y.$$

Here, $H_A$ is the e-ph interaction between the electron and the radiation field, with $\vec{A} = (A_x, A_y)$ being the in-plane vector potential of the field. $H_D^{(0)}$ is the QD$_L$ Hamiltonian in the absence of radiation, with $\Delta(\vec{r})$ and $V(\vec{r})$ being, respectively, the band gap profile and the potential energy profile that give rise to the QD confinement. Let $\phi_D^{(0,c)}$ and $\phi_D^{(0,v)}$ denote the confined states with corresponding energies $E_0^{(c)}$ and $E_0^{(v)}$, respectively. Specifically, they are respectively the lowest conduction band and highest valence band eigenstates of $H_D^{(0)}$ given by

$$\phi_D^{(0,c)} = \begin{cases} \begin{pmatrix} \varphi_{A,>}^{(0,c)} \\ \varphi_{B,<}^{(0,c)} \end{pmatrix}, \tau_v = +1 \\ \begin{pmatrix} \varphi_{A,>}^{(0,c)} \\ -\varphi_{B,<}^{(0,c)} \end{pmatrix}^*, \tau_v = -1 \end{cases}, \phi_D^{(0,v)} = \begin{cases} \begin{pmatrix} \varphi_{A,<}^{(0,v)} \\ \varphi_{B,>}^{(0,v)} \end{pmatrix}, \tau_v = +1 \\ \begin{pmatrix} \varphi_{A,<}^{(0,v)} \\ -\varphi_{B,>}^{(0,v)} \end{pmatrix}^*, \tau_v = -1 \end{cases} \quad (2)$$

where the subscripts "A" and "B" denote the two atomic sites in a graphene unit cell. We take the above QD ground states to be near band edges, and therefore $\varphi_{B,<}^{(0,c)} \approx v_F \hat{p}_+ \varphi_{A,>}^{(0,c)} / 2\Delta \ll \|\varphi_{A,>}^{(0,c)}\|$ and $\varphi_{A,<}^{(0,v)} \approx -v_F \hat{p}_- \varphi_{B,>}^{(0,v)} / 2\Delta \ll \|\varphi_{B,>}^{(0,v)}\|$. The states with $\tau_v = 1$ and $\tau_v = -1$ are related by time reversal symmetry and are basically complex conjugates of each other as given above. Near resonance $(\omega_{ph} \approx E_0^{(c)} - E_0^{(v)})$, the optical response is governed by the optical matrix element $M \equiv \langle \phi_D^{(0,c)} | H_A | \phi_D^{(0,v)} \rangle$. Below we analyze $M$, with the radiation field representing either a DBR or PC cavity photon.

We start with the DBR cavity mode. In typical applications the mode has a wave length much greater than the size of the qubit QDs. Therefore, we take its electric field inside QD$_L$ to be approximately constant, with the in-plane component $\vec{E}_0 = \vec{E}_{DC}(\vec{r} = 0, z = 0)$ and the corresponding in-plane vector potential $\vec{A} = \frac{\vec{E}_0}{i\omega_{ph}} e^{-i\omega_{ph}t}$. Here, we take the graphene plane to be located at $z = 0$, and QD$_L$ at $\vec{r} \equiv (x, y) = 0$ of the plane. Below we list the matrix elements for various combinations of $\vec{E}_0$ polarization and valley pseudospin ($E_0 = |\vec{E}_0|$):

$$\begin{cases} M_> = ev_F \left\langle \varphi_{A,>}^{(0,c)} \middle| -\frac{iE_0}{\omega_{ph}} \middle| \varphi_{B,>}^{(0,v)} \right\rangle \text{ for } (\sigma_+, K) \\ M_< = ev_F \left\langle \varphi_{B,<}^{(0,c)} \middle| -\frac{iE_0}{\omega_{ph}} \middle| \varphi_{A,<}^{(0,v)} \right\rangle \text{ for } (\sigma_-, K) \\ -M_<^* = ev_F \left\langle \varphi_{B,<}^{(0,c)*} \middle| -\frac{iE_0}{\omega_{ph}} \middle| \varphi_{A,<}^{(0,v)*} \right\rangle \text{ for } (\sigma_+, K') \\ -M_>^* = ev_F \left\langle \varphi_{A,>}^{(0,c)*} \middle| -\frac{iE_0}{\omega_{ph}} \middle| \varphi_{B,>}^{(0,v)*} \right\rangle \text{ for } (\sigma_-, K') \end{cases} \quad (3)$$

In the above, ($\sigma_+$, K) denotes the absorption of a $\sigma_+$ polarized photon by a K valence electron in QD$_L$, for example. Basically, for near band edge states, because $|\phi_>| \gg |\phi_<|$, it gives $|M_>| \gg |M_<|$. If we totally ignore the minor matrix element $M_<$, then Eqn. (3) yields, for optical excitation, the major matrix element $M_>$ consistent with the valley-polarization correspondence in **Figure 1**. However, since $M_<$ is finite, the correspondence is only approximate. After substituting $\varphi_{B,<}^{(0,c)} \approx v_F \hat{p}_+ \varphi_{A,>}^{(0,c)} / 2\Delta$ and $\varphi_{A,<}^{(0,v)} \approx -v_F \hat{p}_- \varphi_{B,>}^{(0,v)} / 2\Delta$ into $M_<$, we obtain

$$M_< \approx ev_F \left\langle \frac{1}{2\Delta} v_F \hat{p}_+ \varphi_{A,>}^{(0,c)} \middle| -\frac{iE_0}{\omega_{ph}} \middle| \frac{1}{-2\Delta} v_F \hat{p}_- \varphi_{B,>}^{(0,v)} \right\rangle$$
$$\text{for } (\sigma_-, K)$$
$$-M_<^* \approx ev_F \left\langle \frac{1}{2\Delta} v_F \hat{p}_- \varphi_{A,>}^{(0,c)*} \middle| -\frac{iE_0}{\omega_{ph}} \middle| \frac{1}{-2\Delta} v_F \hat{p}_+ \varphi_{B,>}^{(0,v)*} \right\rangle \quad (4)$$
$$\text{for } (\sigma_+, K')$$

The above expressions are useful for the estimation of the parameters $M_>$ and $M_<$ in our numerical investigation of the QST. In general, the ratio $M_</M_>$ depends on the QD geometry as well as the confinement. For example, in a QD with electron-hole symmetry, Eqns. (3) and (4) imply that $M_</M_> \propto \left\langle \hat{p}_+ \varphi_{A,>}^{(0,c)} \middle| \hat{p}_- \varphi_{A,>}^{(0,c)} \right\rangle \approx \left\langle k_x^2 - k_y^2 - 2ik_x k_y \right\rangle$. In the case of a circular disk QD, $\left\langle k_x^2 - k_y^2 \right\rangle = <k_x k_y> = 0$, so $M_< = 0$, while in an elliptic QD, $<k_x k_y>$ vanishes and so does the imaginary part of $M_<$. For a generic, asymmetric QD, $<k_x k_y>$ is likely to be finite, so $M_<$ generally carries a phase relative to $M_>$. Last, we note that the above discussion of matrix elements has been performed within the one-electron picture. As such, for trion-involved optical transitions considered in the study, the expressions of matrix elements will be modified by many-electron effects. However, since the primary key to the photon-valley QST - the relations among the various matrix elements in Eqns. (3) and (4) - is established only on the basis of the time reversal symmetry relating the two states of opposite pseudospins, its validity holds even in the presence of many-electron effects, as briefly explained below. Consider the trion in QD$_L$ with one K electron, one K hole, and one K' electron. In the presence of Coulomb interaction, the trion wave function is given by [46]

$$\Psi_{trion}^{(K)} = \sum_{k_e, k_h, k_e'} c(k_e, k_h, k_e') |k_e + K, k_h + K, k_e' + K'\rangle.$$

A similar expression holds for $\Psi_{trion}^{(K')}$. Here, $|k_e + K, k_h + K, k_e' + K'\rangle$ is a basis state consisting of two free electrons and one hole, with corresponding wave vectors as specified in the ket, and $c(k_e, k_h, k_e')$ is the corresponding expansion coefficient. Distinct basis states are coupled together by the electron-electron and electron-hole Coulomb interaction as well as the QD confinement potential, giving the trion state as a linear combination of these basis states. This generalizes the final state in optical absorption considered in Eqn. (3) from a free conduction band electron in one-electron picture to the corresponding trion in many-electron picture. Moreover, let $H_{trion}$ denote the corresponding trion Hamiltonian. Due to the time reversal symmetry between K and K', it follows that [46]

$$H_{trion} \Psi_{trion}^{(K)} = E \Psi_{trion}^{(K)}$$
$$H_{trion} \Psi_{trion}^{(K')} = E \Psi_{trion}^{(K')},$$
$$\Psi_{trion}^{(K')} = (\Psi_{trion}^{(K)})^*,$$

which generalizes the final state identity from $\varphi_{A,>}^{(0,c)} = (\varphi_{A,>}^{(0,c)*})^*$, $\varphi_{B,<}^{(0,c)} = (\varphi_{B,<}^{(0,c)*})^*$ in Eqn. (3) to $\Psi_{trion}^{(K')} = \Psi_{trion}^{K*}$. Since the relations among the various matrix elements in Eqn. (3) depend only on such final state identities, we conclude that in the presence of many-electron effects the same relations continue to hold and many-electron effects only modify the matrix elements in magnitude. This modification in magnitude is covered in our work by giving the matrix elements a range of magnitudes and studying the QST as a function of these magnitudes. Therefore, while one-electron expressions in Eqns. (3) and (4) will be used below for numerical estimation of the matrix elements, by adopting the forgoing approach our study does not depend so much on the validity of one-electron approximation and actually allows the study to go beyond the approximation.

Next, we discuss the optical matrix element involving the PC cavity mode. We take the mode to be a TE donor type state at X point, for example, one that transforms according to the symmetry of two-dimensional E representation of $C_{4v}$, i.e., the symmetry group of a square [47]. There are two degenerate modes in the representation as follows. Let $H_{PC}(\vec{r})\hat{z}$ and $\vec{E}_{PC}(\vec{r})$ denote the H-field and E-field of the modes in the graphene plane, respectively ($\hat{z}$ = unit vector normal to the plane). Then, $H_{PC}(\vec{r})$ transforms as $\sin(\vec{k}_{X_1} \cdot \vec{r})$ or $\sin(\vec{k}_{X_2} \cdot \vec{r})$, and $\vec{E}_{PC}(\vec{r}) = \frac{1}{-i\omega_{ph}\mu_0\varepsilon(\vec{r})} \nabla \times H_{PC}(\vec{r})\hat{z}$ transforms as $\hat{z} \times \vec{k}_{X_1} \cos(\vec{k}_{X_1} \cdot \vec{r})$ or $\hat{z} \times \vec{k}_{X_2} \cos(\vec{k}_{X_2} \cdot \vec{r})$. Here { $\vec{k}_{X_1}$, $\vec{k}_{X_2}$ } are the two orthogonal wave vectors at X points of the Brillouin zone, $\mu_0$ = vacuum magnetic permeability and $\varepsilon$ = dielectric constant. Note that at $\vec{r} = 0$ (center of the cavity) where the qubit is located, $\vec{E}_{PC}(\vec{r})$'s of the two modes are {$\sigma_x$, $\sigma_y$} polarized, in a way correlated to their propagation directions { $\vec{k}_{X_1}$, $\vec{k}_{X_2}$ }. For the optical matrix elements, we can linearly combine the two modes, making it either $\sigma_+$ or $\sigma_-$ polarized at $\vec{r} = 0$, and continue using the same expressions in Eqn. (3) with $\vec{E}_0$ replaced by $\vec{E}_{PC}(\vec{r} = 0)$. Effectively, this means that the matrix elements for the PC cavity are scaled from those for the DBR cavity by the factor $|\vec{E}_{PC}(\vec{r} = 0)/\vec{E}_0|$. In particular, it follows that the major matrix elements for the two cavities carry the same phase and so do the minor ones.

Eqns. (3) and (4) are used to estimate the matrix elements for the numerical study in **Sec. IV**. As an example, we take $\omega_{ph}$ = 1.6·10$^5$ GHz corresponding to a graphene band gap of 0.1eV, and the modal volumes to be $V_{mode}$ = 10$^4$ μm$^3$ for the DBR cavity and $V_{mode}$ = 600 for the PC cavity. As a reference, we also list $(\lambda/n)^3$ = 1600 μm$^3$ for the DBR cavity with the index of refraction "n" taken to be 1 (for air), and $(\lambda/n)^3$ = 94 μm$^3$ for the PC cavity with "n" taken to be 2.6 (in the case of SiC), respectively, where λ is the photon wave length in vacuum. We take the QD to be a square well with edge length of 70 nm and subject to hard wall confinement. This gives the electron velocity v = 0.4v$_F$ in the QD, where v$_F$ (the Fermi velocity) is taken to be 10$^6$ m/sec. Using the above numbers along with the approximations $<\varphi_{A,>}^{(0,c)}|E_0|\varphi_{B,>}^{(0,v)}> \approx E_0$ and $E_0 \sim \sqrt{\hbar/4\pi\varepsilon\omega_{ph}V_{mode}}$ in Eqn. (3), we obtain |M$_>$| ~ 30 GHz for the DBR cavity and 45 GHz for the PC cavity. Moreover, using the approximation $|M_<| \sim \frac{v^2}{4v_F^2}|M_>|$ for Eqn. (4), we obtain |M$_<$/M$_>$| = 0.04. The ratio holds for both cavities since the matrix elements in the two cases are given by the same forms of expressions in Eqns. (3) and (4).

In the following discussion, we introduce the notations {*A*, *B*} and {*C*, *D*} to represent {*M$_>$*, *M$_<$*} for the PC and DBR cavity modes, respectively, with *B/A* = *D/C*. In addition, as it will become obvious below in **II-4**, only the relative phase between *M$_>$* and *M$_<$* matters in the QST, so we take *M$_>$* (*A* and *C*) to be real numbers and place the relative phase in *M$_<$* (*B* and *D*), which allows us to write, for example $B = A|B/A|e^{i\phi_{B/A}}$. Generally, in a favorable configuration design for the QST, a correlation between *A* and *C* would be required so that the design can move the QST process forward to the finish line with a good yield. More details will be given in the following sections.

### II-4. QST with Two Cavities

We examine the QST in the setup with two cavities. In the presence of finite *B* and *D*, the various states involved in the QST are modified from those in **Sec. II-1**, giving

$$|\Phi_0\rangle = \frac{1}{\sqrt{2}}(|K_L K'_R\rangle - |K'_L K_R\rangle) \otimes (\alpha|\sigma_+\rangle + \beta|\sigma_-\rangle),$$

$$|\Phi_1\rangle \propto (-\beta C^* + \alpha D)|K'_{ex,L} K_L K'_R\rangle$$
$$- (\alpha C - \beta D^*)|K_{ex,L} K'_L K_R\rangle,$$

$$|\Phi_2\rangle = \text{valley-polarization entangled state}$$
$$\text{after photon emission from the trion}$$
$$\propto (-\beta C^* + \alpha D)$$
$$(-A|\sigma_-\rangle \otimes |K_L K'_R\rangle + B^*|\sigma_+\rangle \otimes |K_L K'_R\rangle) \quad (5)$$
$$- (\alpha C - \beta D^*)$$
$$(A^*|\sigma_+\rangle \otimes |K'_L K_R\rangle - B|\sigma_-\rangle \otimes |K'_L K_R\rangle),$$

$$|\Phi_{3x}\rangle \propto (-\beta C^* + \alpha D)(-A + B^*)|K_L K'_R\rangle$$
$$- (\alpha C - \beta D^*)(A^* - B)|K'_L K_R\rangle,$$

$$|\Phi_{3y}\rangle \propto (\beta C^* - \alpha D)(A + B^*)|K_L K'_R\rangle$$
$$+ (\alpha C - \beta D^*)(A^* + B)|K'_L K_R\rangle.$$

We note several points. First, the emergence of matrix elements *A*, *B*, *C*, and *D* in Eqn. (5) indicates the presence of two cavities. Second, with two cavities, the photon in the entangled state $|\Phi_2\rangle$ leaks out of the PC cavity via TE modes. For these modes, as their polarizations {$\sigma_x$, $\sigma_y$} are correlated to the propagation directions { $\vec{k}_{X_1}$, $\vec{k}_{X_2}$ }, a detection of the photon's outgoing direction effects the projection of $|\Phi_2\rangle$ onto $|\Phi_{3x}\rangle$ or $|\Phi_{3y}\rangle$. Last, if we set the minor optical matrix elements $B = D = 0$, then the final states are given by $|\Phi_{3x}\rangle \propto \beta|K_L K'_R\rangle - \alpha|K'_L K_R\rangle$ or $|\Phi_{3y}\rangle \propto \beta|K_L K'_R\rangle + \alpha|K'_L K_R\rangle$ with the same amplitudes $\alpha$ and $\beta$ that are encoded into the incoming photon, meaning that no distortion occurs in the QST process. We thus define the entangled state with $B = D = 0$ as

$$|\Psi_{ideal}\rangle \equiv \frac{1}{\sqrt{2}}[(\beta|K_L K'_R\rangle - \alpha|K'_L K_R\rangle) \otimes |\sigma_x\rangle \\ - i(\beta|K_L K'_R\rangle + \alpha|K'_L K_R\rangle) \otimes |\sigma_y\rangle], \quad (6)$$

which serves as a reference state for the definition of fidelity. For example, if we ignore any cavity leakage and qubit decoherence, the fidelity would then be given by $F = N_2 |\langle \Psi_{ideal} | \Phi_2 \rangle|$, with a value less than unity since small yet finite $B$ and $D$ would create in $|\Phi_2\rangle$ a deviation from $|\Psi_{ideal}\rangle$ ($N_2$ = the normalization constant of $|\Phi_2\rangle$). **Sec. III** treats the general case where both the cavity leakage and intermediate state damping are present.

## III. THEORETICAL MODEL

In a realistic system, the QST depends on various parameters of the configuration in which the QST takes place, such as optical transition matrix elements, various decoherence times, and $Q$ factors of cavities. In **Sec. III-1**, we provide a quantum-mechanical description of the realistic QST problem. In **Sec. III-2**, we discuss the wave equation and solution, and derive the yield and fidelity.

### III-1. Description of the Problem

For a quantum-mechanical description of QST in the setup of **Figure 3**, we refer to the following process flow diagram:

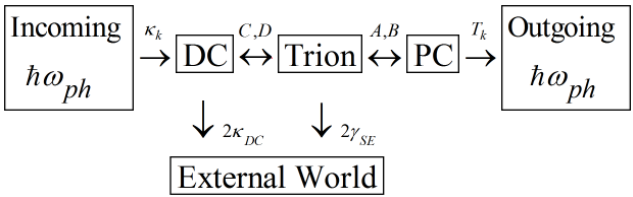

**Figure 4** The process flow diagram. DC denotes the DBR cavity mode and PC denotes the PC cavity mode. The various parameters shown in the diagram are the couplings and decay or leakage rates involved in the flow from one stage to the next, as explained in the text.

In **Figure 4**, the incoming signal photon injects into the DBR cavity with the tunneling coupling given by $\kappa_k$. Next, the cavity photon is absorbed exciting a trion in QD$_L$, with the major (minor) coupling given by $C$ ($D$) between the trion and the DBR cavity mode. The trion then radiates a photon into the PC cavity mode, with the major (minor) coupling given by $A$ ($B$) between the trion and the PC cavity mode. Last, the photon leaks out of the PC cavity with the coupling constant given by $T_k$, and enters a photon sensor. The parameter "$2\kappa_{DC}$" is the leakage rate of DBR cavity mode, and "$2\gamma_{SE}$" the decay rate of trion accounting for its nonradiative decay as well as emission into modes excluding the DBR and PC cavity ones. Note that we assume the coupling between the DRB and PC cavities is negligible because of the significant mode mismatch between them.

The quantum-mechanical system involved consists of electrons, trions, cavity photons, incoming and outgoing photons, and the interaction among them. Below we discuss the state vector, Hamiltonian, wave equation, and last, the solution to the equation.

**(State Vector)** The total state vector describes an e-ph composite system and is given by

$$|\Psi(t)\rangle = \sum_{\sigma=\sigma_+,\sigma_-;\tau=K,K'} \int dx\, \phi^{input}_{\sigma\tau}(x,t)|\tau'_L,\tau_R\rangle \otimes |x,\sigma\rangle \\
+ \sum_{\sigma=\sigma_+,\sigma_-;\tau=K,K'} \phi^{DC}_{\sigma\tau}(t)|\tau'_L,\tau_R\rangle \otimes |DC,\sigma\rangle \\
+ \sum_{\tau=K,K'} \phi^{trion}_{\tau}(t)|\tau_{ex,L},\tau'_L,\tau_R\rangle \quad (7) \\
+ \sum_{\sigma=\sigma_+,\sigma_-;\tau=K,K'} \phi^{PC}_{\sigma\tau}(t)|\tau'_L,\tau_R\rangle \otimes |PC,\sigma\rangle \\
+ \sum_{\sigma=\sigma_+,\sigma_-;\tau=K,K';k_{2D}} \phi^{output}_{\sigma\tau}(k_{2D},t)|\tau'_L,\tau_R\rangle \otimes |k_{2D},\sigma\rangle$$

In the above expression, we define $\tau' = K(K')$ for $\tau = K'(K)$. $|\tau'_L,\tau_R\rangle$ denotes the two-electron state with the QD$_L$ electron in the $\tau'_L$ valley and the QD$_R$ electron in the opposite $\tau_R$ valley; $|x,\sigma\rangle$ denotes the incoming signal photon state with position $x$ and circular polarization $\sigma$; $|DC,\sigma\rangle$ denotes a DBR cavity mode with polarization $\sigma$ at the qubit; $|\tau_{ex,L},\tau'_L,\tau_R\rangle$ denotes a trion-electron state, with the trion (specified by $\tau_{ex,L},\tau'_L$) in QD$_L$ and the electron (specified by $\tau_R$) in QD$_R$; $|PC,\sigma\rangle$ denotes a PC cavity mode with polarization $\sigma$ at the qubit; and $|k_{2D},\sigma\rangle$ denotes an outgoing signal photon state moving to the photon sensor with a planar wave vector $k_{2D}$ and polarization $\sigma$. $\phi^{input}_{\sigma\tau}$, $\phi^{DC}_{\sigma\tau}$, $\phi^{trion}_{\tau}$, $\phi^{PC}_{\sigma\tau}$, and $\phi^{output}_{\sigma\tau}$ are the amplitudes of various basis states and governed by the corresponding wave equation (See **Sec. III-2**). $\phi^{input}_{\sigma\tau}$ is used as an input to the equation. The fourteen amplitudes $\{\phi^{DC}_{\sigma\tau}, \phi^{trion}_{\tau}, \phi^{PC}_{\sigma\tau}, \phi^{output}_{\sigma\tau}$ for $\sigma = \sigma_+,\sigma_-$ and $\tau = K,K'\}$ are determined by solving the equation. $\phi^{output}_{\sigma\tau}$ describes the output from the system and, once solved, it is used to determine both the yield and fidelity.

**(Input to the system)** The e-ph system is treated as an open system with both input from and output to the external world. In particular, in the corresponding wave equation, the input is prescribed in advance as a time-dependent boundary condition. In details, the incoming photonic signal is taken to be a Gaussian wave packet and, together with the initial singlet state of the valley qubit, it leads to the following product state as the input to the e-ph system:

$$|\Phi^{input}\rangle = G(t)(\alpha|\sigma_+\rangle + \beta|\sigma_-\rangle) \otimes \frac{1}{\sqrt{2}}(|K_L K_R'\rangle - |K_L' K_R\rangle), \quad (8)$$

where $G(t) \propto \phi_{\sigma\tau}^{input}(x=0,t)$ denotes the Gaussian wave packet evaluated at the DBR cavity-environment interface the incoming photon hits upon, with the interface taken to be located at x = 0. Specifically, it is given by

$$G(t) = \int_{-\infty}^{\infty} \frac{d\omega}{2\pi} \phi_0^{ph} \sqrt{L} e^{-i(\omega-\omega_{ph})x_0/c} e^{-(\omega-\omega_{ph})^2/2\Delta\omega_{ph}^2} e^{-i\omega t},$$

$$\phi_0^{ph} \equiv \pi^{\frac{1}{4}} \sqrt{\frac{2}{c\Delta\omega_{ph}}}, \quad (9)$$

with $x_0$ the initial center of the wave packet at t = 0, $\Delta\omega_{ph}$ the band width, L the size of the external world, and c the speed of light. The wave packet above has been taken to be primarily composed of waves normally incident upon the DBR. Expanding the product, we obtain the four components of $|\Phi^{input}\rangle$ as

$$|\phi^{input}\rangle = \left(\phi_{\sigma_+,K}^{input}(t), \phi_{\sigma_-,K}^{input}(t), \phi_{\sigma_+,K'}^{input}(t), \phi_{\sigma_-,K'}^{input}(t)\right)^T \quad (10)$$
$$= 1/\sqrt{2}\left(-\alpha G(t), -\beta G(t), \alpha G(t), \beta G(t)\right)^T$$

which are inputs to the wave equation discussed later.

**(Hamiltonian)** The Hamiltonian of the e-ph system is given by

$$H = H_{input} + H_{DC} + H_{trion} + H_{PC} + H_{output} + H_{reservoir}$$
$$+ H_{input-DC} + H_{DC-trion} + H_{trion-PC} + H_{PC-output} \quad (11)$$
$$+ H_{SE}.$$

The various terms are given as follows ($\hbar = 1$):

$$H_{input} = \sum_{\substack{\sigma=\sigma_+,\sigma_-;\\ \tau=K,K';\\ k_{1D}}} c|k_{1D}||\tau_L'\tau_R\rangle \otimes |k_{1D},\sigma\rangle\langle k_{1D},\sigma| \otimes \langle\tau_L'\tau_R|,$$

$$H_{DC} = \sum_{\substack{\sigma=\sigma_+,\sigma_-;\tau=K,K'}} \omega_{DC}|\tau_L'\tau_R\rangle \otimes |DC,\sigma\rangle\langle DC,\sigma| \otimes \langle\tau_L'\tau_R|,$$

$$H_{trion} = \sum_{\sigma=\sigma_+,\sigma_-;\tau=K,K'} \omega_{trion}|\tau_{ex,L}\tau_L'\tau_R\rangle\langle\tau_{ex,L}\tau_L'\tau_R|,$$

$$H_{PC} = \sum_{\sigma=\sigma_+,\sigma_-;\tau=K,K'} \omega_{PC}|\tau_L'\tau_R\rangle \otimes |PC,\sigma\rangle\langle\tau_L'\tau_R| \otimes \langle PC,\sigma|,$$

$$H_{output} = \sum_{\substack{\sigma=\sigma_+,\sigma_-;\\ \tau=K,K';\\ \vec{k}_{2D}}} c|\vec{k}_{2D}||\tau_L'\tau_R\rangle \otimes |\vec{k}_{2D},\sigma\rangle\langle\tau_L'\tau_R| \otimes \langle\vec{k}_{2D},\sigma|,$$

$$H_{reservoir} = \sum_{\sigma=\sigma_+,\sigma_-;\tau=K,K';\mu} \omega_\mu |\tau_L'\tau_R\rangle \otimes |\mu,\sigma\rangle\langle\mu,\sigma| \otimes \langle\tau_L'\tau_R|,$$

$$H_{input-DC} = \sum_{\substack{\sigma=\sigma_+,\sigma_-;\\ \tau=K,K';\\ k_{1D}}} \left(\kappa_k|\tau_L'\tau_R\rangle \otimes |DC,\sigma\rangle\langle\tau_L'\tau_R| \otimes \langle k_{1D},\sigma| + h.c.\right),$$

$$\begin{aligned}H_{trion-PC} &= A|K_{ex,L}K_L'K_R\rangle\langle K_L'K_R| \otimes \langle PC,\sigma_+|\\ &+ B|K_{ex,L}K_L K_R\rangle\langle K_L K_R| \otimes \langle PC,\sigma_-|\\ &- B^*|K_{ex,L}'K_L K_R'\rangle\langle K_L K_R'| \otimes \langle PC,\sigma_+|\\ &- A^*|K_{ex,L}'K_L K_R'\rangle\langle K_L K_R'| \otimes \langle PC,\sigma_-|\\ &+ h.c.,\end{aligned}$$

$$H_{PC-output} = \sum_{\substack{\sigma=\sigma_+,\sigma_-;\\ \tau=K,K';\\ \vec{k}_{2D}}} \left(T_k|\tau_L'\tau_R\rangle \otimes |\vec{k}_{2D},\sigma\rangle\langle\tau_L'\tau_R| \otimes \langle PC,\sigma| + h.c.\right),$$

$$H_{SE} = \sum_{\sigma=\sigma_+,\sigma_-;\tau=K,K';\mu} \left(\gamma_\mu|\tau_L'\tau_R\rangle|\mu,\sigma\rangle\langle\tau_{ex,L}\tau_L'\tau_R| + h.c.\right). \quad (12)$$

The above Hamiltonian includes contributions from several subsystems. $H_{input}$ comes from the photon states that are outside the DBR cavity and propagate along the cavity axis, with $k_{1D}$ the wave vector of the photon; $H_{DC}$ comes from the DBR cavity modes with $\omega_{DC}$ the mode frequency; $H_{trion}$ comes from the trion states with $\omega_{trion}$ the trion frequency (relative to that of the qubit electron in QD$_L$); $H_{PC}$ comes from the PC cavity modes with $\omega_{PC}$ the mode frequency; $H_{output}$ comes from the photon states that are outside the PC cavity and propagate in the plane, with $k_{2D}$ being the wave vector of the photon; and $H_{reservoir}$ comes from the photon reservoir, excluding the contributions $H_{input}$, $H_{DC}$, $H_{PC}$, and $H_{output}$ ($\mu$ = photon state label and $\omega_\mu$ = photon frequency). Throughout the work, we assume the resonance condition $\omega_{ph} = \omega_{DC} = \omega_{PC} = \omega_{trion}$. The Hamiltonian also includes couplings among the subsystems. $H_{input-DC}$ describes the tunneling of photons into and out of the DBR cavity; $H_{DC-trion}$ describes the coupling between the DBR cavity mode and the trion; $H_{trion-PC}$ describes the coupling between the trion and the PC cavity mode; $H_{PC-output}$

describes the tunneling of photons into and out of the PC cavity; and $H_{SE}$ describes the coupling between the trion and the photon reservoir with the coupling constant $\gamma_\mu$. Several approximations have been made above or will be made below. For example, we take $\kappa_k \approx \sqrt{c\kappa_{DC}/L}$ ($T_k \approx \sqrt{2c^2 \Gamma_{PC}/L^2 \omega_{PC}}$) independent of the wave number under the flat-band assumption, with $2\kappa_{DC}$ ($2\Gamma_{PC}$) = leakage rate of the DBR cavity (PC cavity) mode. Equivalently, this means that we ignore the leakage into the reservoir and take the leakage rate to be primarily due to the coupling $H_{input\text{-}DC}$ ($H_{PC\text{-}output}$), with $\kappa_{DC} \approx \pi \sum_{k_{1D}} |\kappa_k|^2 \delta(\omega_{DC} - c|k_{1D}|)$ ($\Gamma_{PC} \approx \pi \sum_{\vec{k}_{2D}} |T_k|^2 \delta(\omega_{PC} - c|\vec{k}_{2D}|)$). Moreover, in Eqn. (12) we have neglected the exchange process $|K_{ex,L}, K'_L\rangle \leftrightarrow |K'_{ex,L}, K_L\rangle$ where the simultaneous valley flips of three carriers are involved. As such a process is of high order, it is neglected in the equation. This approximation decouples the amplitudes of the trion states $|K_{ex,L}, K'_L\rangle$ and $|K'_{ex,L}, K_L\rangle$ and, thus, facilitates the solution to the wave equation, as will become clear below. Last, the rate of trion emission into the reservoir is given by $2\gamma_{SE}(\omega) = 2\pi \sum_\mu |\gamma_\mu|^2 \delta(\omega - \omega_\mu)$. We take $\gamma_{SE}(\omega)$ as a phenomenological constant that also accounts for nonradiative decay of the trion.

### III-2. Wave Equation and Solution

Using the Hamiltonian specified above, we set up the wave equation for the system. For typical applications, as the cavity leakage rates $\Gamma_{PC}$ and $\kappa_{DC}$ scale inversely with the corresponding cavity Q factors, we take them to be the maximum frequency parameters so as not to impose stringent requirements on the Q factors. Moreover, we take $|B/A| \ll 1$, according to the numerical estimate obtained in **Sec. II-3** for a typical QD.

The wave equation consists of coupled differential equations for the fourteen amplitudes { $\phi^{DC}_{\sigma\tau}$, $\phi^{trion}_{\tau}$, $\phi^{PC}_{\sigma\tau}$, $\phi^{output}_{\sigma\tau}$ }, which are divided into three sets and approximately solved. The three sets of equations describe the key sub-processes in the QST, respectively, as follows.

The first set of equations govern $\phi^{DC}_{\sigma\tau}$ and the process "incident signal photon → DBR cavity photon". They are given by

$$i\partial_t \phi^{DC}_{\sigma\tau}(t) \approx (\omega_{DC} - i\kappa_{DC})\phi^{DC}_{\sigma\tau}(t) + \sqrt{2c\kappa_{DC}/L}\phi^{input}_{\sigma\tau}(t). \quad (13)$$

Eqn. (13) takes $\phi^{DC}_{\sigma\tau}$ to be primarily determined by the incoming signal $\phi^{input}_{\sigma\tau}$ and the damping ($\kappa_{DC}$) of DBR cavity modes. For simplification, it neglects the contributions from the process of photon emission by the trion or that of photon absorption by the qubit, both of which occur only after the entry of signal photon into the cavity as indicated in **Figure 4** and, hence, are higher-order processes from the perturbation-theoretical point of view. After solving Eqn. (13), $\phi^{DC}_{\sigma\tau}$ is given in terms of the input by

$$\begin{aligned}
&\phi^{DC}_{\sigma\tau}(t) \\
&= i\bar{\alpha}(\sigma,\tau)\sqrt{c\kappa_{DC}/L}\int_0^t dt' e^{-i(\omega_{DC}-i\kappa_{DC})(t-t')} G(t'), \\
&\bar{\alpha}(\sigma_+, K) = \alpha, \bar{\alpha}(\sigma_+, K') = -\alpha, \bar{\alpha}(\sigma_+, K') = \beta, \bar{\alpha}(\sigma_-, K') = -\beta
\end{aligned} \quad (14)$$

The second set of equations govern $\phi^{trion}_\tau$, $\phi^{PC}_{\sigma\tau}$ and the resonant process of "photon + electron ↔ trion". For the specific process |photon, $K_L$⟩ ↔ |$K'_{ex}$, $K_L$⟩, for example, we obtain

$$i\partial_t \begin{pmatrix} \phi^{tri}_K(t) \\ \phi^{PC}_{\sigma_+ K}(t) \\ \phi^{PC}_{\sigma_- K}(t) \end{pmatrix}
= \begin{pmatrix} (\omega_{trion} - i\gamma_{total}) & A & B \\ A^* & (\omega_{PC} - i\Gamma_{PC}) & 0 \\ B^* & 0 & (\omega_{PC} - i\Gamma_{PC}) \end{pmatrix} \begin{pmatrix} \phi^{tri}_K(t) \\ \phi^{PC}_{\sigma_+ K}(t) \\ \phi^{PC}_{\sigma_- K}(t) \end{pmatrix}$$
$$+ \begin{pmatrix} C\phi^{DC}_{\sigma_+ K}(t) + D\phi^{DC}_{\sigma_- K}(t) \\ 0 \\ 0 \end{pmatrix} \quad (15)$$

$\gamma_{total} = \gamma_{SE} + \gamma_{tD}$.

$\gamma_{SE}$, $\gamma_{tD}$, and $\Gamma_{PC}$ account for trion and PC cavity photon decays due to the couplings $H_{SE}$, $H_{DC\text{-}trion}$, and $H_{PC\text{-}output}$, respectively. $\gamma_{tD} \approx \pi |C|^2 D_{os} = |C|^2/\kappa_{DC}$, where $D_{os} = 1/\pi\kappa_{DC}$ is the density of states for the DBR cavity mode with level broadening due to the cavity leakage. Note that the DBR cavity photon, with the amplitude $\phi^{DC}_{\sigma\tau}$ determined by Eqn. (14), now provides a source term to Eqn. (15) feeding photons into the resonant process. For the other process |photon, $K'_L$⟩ ↔ |$K_{ex}$, $K'_L$⟩, a similar set of equations are obtained by an appropriate change of valley and polarization subscripts in Eqn. (15).

Last, the 3$^{rd}$ set of equations governs $\phi^{output}_{\sigma\tau}$ and describes the process "PC cavity photon → outgoing photon":

$$i\partial_t \phi^{output}_{\sigma\tau}(k_{2D}, t) = T_k \phi^{PC}_{\sigma\tau}(t) + \omega_{output} \phi^{output}_{\sigma\tau}(t) \quad (16)$$

where $\omega_{output} = c|\vec{k}_{2D}|$. The argument $k_{2D}$ in $\phi^{output}_{\sigma\tau}$ will be omitted below when it does not cause confusion.

In order to solve Eqns. (15) and (16), we perform a linear transformation and obtain

$$i\partial_t \begin{pmatrix} \phi_K^{trion}(t) \\ \phi_{K1}^{PC}(t) \\ \phi_{K2}^{PC}(t) \end{pmatrix}$$

$$= \begin{pmatrix} (\omega_{trion} - i\gamma_{total}) & \sqrt{|A|^2 + |B|^2} & 0 \\ \sqrt{|A|^2 + |B|^2} & (\omega_{PC} - i\Gamma_{PC}) & 0 \\ 0 & 0 & (\omega_{PC} - i\Gamma_{PC}) \end{pmatrix} \begin{pmatrix} \phi_K^{trion}(t) \\ \phi_{K1}^{PC}(t) \\ \phi_{K2}^{PC}(t) \end{pmatrix}$$

$$+ \begin{pmatrix} C\phi_{\sigma_+ K}^{DC}(t) + D\phi_{\sigma_- K}^{DC}(t) \\ 0 \\ 0 \end{pmatrix} \quad (17)$$

and

$$i\partial_t \begin{pmatrix} \phi_{K1}^{output}(t) \\ \phi_{K2}^{output}(t) \end{pmatrix} = T_k \begin{pmatrix} \phi_{K1}^{PC}(t) \\ \phi_{K2}^{PC}(t) \end{pmatrix} + \omega_{output} \begin{pmatrix} \phi_{K1}^{output}(t) \\ \phi_{K2}^{output}(t) \end{pmatrix} \quad (18)$$

with the transformation given by

$$\phi_{K1}^{PC}(t) = \frac{A}{\sqrt{|A|^2 + |B|^2}} \phi_{\sigma_+ K}^{PC}(t) + \frac{B}{\sqrt{|A|^2 + |B|^2}} \phi_{\sigma_- K}^{PC}(t)$$

$$\phi_{K2}^{PC}(t) = -\frac{B^*}{\sqrt{|A|^2 + |B|^2}} \phi_{\sigma_+ K}^{PC}(t) + \frac{A^*}{\sqrt{|A|^2 + |B|^2}} \phi_{\sigma_- K}^{PC}(t) \quad (19)$$

and

$$\phi_{K1}^{output}(t) = \frac{A}{\sqrt{|A|^2 + |B|^2}} \phi_{\sigma_+ K}^{output}(t) + \frac{B}{\sqrt{|A|^2 + |B|^2}} \phi_{\sigma_- K}^{output}(t)$$

$$\phi_{K2}^{output}(t) = -\frac{B^*}{\sqrt{|A|^2 + |B|^2}} \phi_{\sigma_+ K}^{output}(t) + \frac{A^*}{\sqrt{|A|^2 + |B|^2}} \phi_{\sigma_- K}^{output}(t) \quad (20)$$

The transformed equations can be solved as follows. First, the block of top two rows in Eqn. (17) forms a two-component time-dependent Schrodinger equation with a source term:

$$i\partial_t \begin{pmatrix} \phi_K^{trion}(t) \\ \phi_{K1}^{PC}(t) \end{pmatrix} = H_0 \begin{pmatrix} \phi_K^{trion}(t) \\ \phi_{K1}^{PC}(t) \end{pmatrix} + |f(t)\rangle, \quad (21)$$

where

$$H_0 = \begin{pmatrix} a & \sqrt{|A|^2 + |B|^2} \\ \sqrt{|A|^2 + |B|^2} & b \end{pmatrix},$$

$$a \equiv \omega_{trion} - i\gamma_{total}, \quad b \equiv \omega_{PC} - i\Gamma_{PC},$$

$$|f(t)\rangle \equiv \begin{pmatrix} C\phi_{\sigma_+ K}^{DC}(t) + D\phi_{\sigma_- K}^{DC}(t) \\ 0 \end{pmatrix}. \quad (22)$$

The solution to Eqn. (21) with the initial condition $\phi_K^{trion}(0) = \phi_{K1}^{PC}(0) = 0$ is given by

$$\begin{pmatrix} \phi_K^{trion}(t) \\ \phi_{K1}^{PC}(t) \end{pmatrix} = -i \sum_{n=1,2} |\varphi_n\rangle \int_0^t e^{-i\lambda_n(t-t')} (\varphi_n | f(t')) dt' \quad (23)$$

where $|\varphi_n\rangle$'s and $\lambda_n$'s for $n = 1, 2$ are, respectively, the eigenvectors and eigenvalues of $H_0$ given in the **Appendix**. Using the expressions there, one can show that $\lambda_1 \approx \omega_{ph} + O(\Gamma_{PC})$ and $\lambda_2 \approx \omega_{ph} + O(\max(\gamma_{total}, \gamma_{tp}))$ under the resonance condition, with $\gamma_{tp} \equiv |A|^2 / \Gamma_{PC}$ the rate of photon emission by the trion into the PC cavity mode. $(\varphi_n | f(t'))$ in Eqn. (23) denotes the projection of $|f(t')\rangle$ onto $|\varphi_n\rangle$ in the case where $H_0$ is non-Hermitian (See **Appendix**). Last, the third row of Eqn. (17) can easily be solved. With initial condition $\phi_{K2}^{PC}(0) = 0$, it leads to $\phi_{K2}^{PC}(t) = 0$.

**(Output from the system)** Eqn. (18) gives the outgoing photon amplitude

$$\phi_{K1}^{output}(t) = -iT_k \int_0^t \phi_{K1}^{PC}(t') e^{-i\omega_{output}(t-t')} dt' \quad (24)$$

and a similar expression for $\phi_{K2}^{output}(t)$. By substituting $\phi_{K1}^{PC}(t)$ and $\phi_{K2}^{PC}(t)$ obtained above, we find the amplitudes at the completion of QST, with the corresponding probabilities given by (See **Appendix**)

$$\lim_{x_0 \to -\infty} \left| \phi_{K1}^{output}(k_{2D}, t \to \infty) \right|^2$$

$$= 2\sqrt{\pi} \frac{\kappa_{DC}}{\Delta\omega_{ph}} e^{-(\omega_{output}(k_{2D}) - \omega_{ph})^2 / \Delta\omega_{ph}^2}$$

$$\frac{|T_k|^2 |\alpha C + \beta D|^2 (|A|^2 + |B|^2)}{\left| (\omega_{output}(k_{2D}) - \lambda_1)(\omega_{output}(k_{2D}) - \lambda_2)(\omega_{output}(k_{2D}) - \lambda_3) \right|^2},$$

$$\left| \phi_{K2}^{output}(k_{2D}, t \to \infty) \right|^2 = 0, \quad (25)$$

$$\lambda_3 \equiv \omega_{DC} - i\kappa_{DC}.$$

A similar procedure obtains $\left| \phi_{K'1}^{output}(k_{2D}, \infty) \right|^2$ and $\left| \phi_{K'2}^{output}(k_{2D}, \infty) \right|^2$, with $|\alpha C + \beta D|^2$ in $\left| \phi_{K1}^{output}(k_{2D}, \infty) \right|^2$ replaced by $|\alpha D^* + \beta C^*|^2$, and $\left| \phi_{K'2}^{output}(k_{2D}, \infty) \right|^2 = 0$.

**(Yield)** Using the forgoing solutions, we derive the figure of merits for the photon-valley QST. First, the yield is given by integrating the various output amplitudes as follows

$$P = \sum_{\sigma,\tau,k_{2D}} \left| \phi_{\sigma\tau}^{output}(k_{2D}, \infty) \right|^2 = \sum_{k_{2D}} P_{k_{2D}},$$

$$P_{k_{2D}} \equiv \left| \phi_{K1}^{output}(k_{2D}, \infty) \right|^2 + \left| \phi_{K'1}^{output}(k_{2D}, \infty) \right|^2, \quad (26)$$

with $P_{k_{2D}}$ the yield for a given $k_{2D}$. This gives

$$P = \frac{2}{\sqrt{\pi}} \frac{\kappa_{DC}\Gamma_{PC}}{\Delta\omega_{ph}} \Big[|\alpha C + \beta D|^2 + |\alpha D^* + \beta C^*|^2\Big]\Big(|A|^2 + |B|^2\Big) I_w,$$

$$I_w \equiv \int d\omega_{output} \frac{e^{-(\omega_{output}-\omega_{ph})^2/\Delta\omega_{ph}^2}}{\left|(\omega_{output}-\lambda_1)(\omega_{output}-\lambda_2)(\omega_{output}-\lambda_3)\right|^2}. \quad (27)$$

In order to gain insights into the dependence of $P$ on various parameters, we analyze the frequency integral $I_w$ in the following. In essence, it is primarily determined by various frequency parameters, such as the photon bandwidth $\Delta\omega_{ph}$ of the Gaussian function $e^{-(\omega_{output}(k_{2D})-\omega_{ph})^2/\Delta\omega_{ph}^2}$ and the poles $\lambda_1$, $\lambda_2$, and $\lambda_3$ in the integrand. In accordance with the condition given earlier that $\Gamma_{PC}$ and $\kappa_{DC}$ are taken to be the maximum frequency parameters, we consider the two following cases, namely, Case 1 where $\min(\Gamma_{PC},\kappa_{DC}) \geq \max(\gamma_{total},\gamma_{tP}) \geq \Delta\omega_{ph}$ and Case 2 where $\min(\Gamma_{PC},\kappa_{DC}) \geq \Delta\omega_{ph} \geq \max(\gamma_{total},\gamma_{tP})$. Below we express $P$ in terms of the dimensionless frequencies $\gamma'_{tD} \equiv \gamma_{tD}/\gamma_{tP}$, $\gamma'_{SE} \equiv \gamma_{SE}/\gamma_{tP}$, and $\Delta\omega'_{ph} \equiv \Delta\omega_{ph}/\gamma_{tP}$, in the two cases.

**Case 1.** For $\min(\Gamma_{PC},\kappa_{DC}) \geq \max(\gamma_{total},\gamma_{tP}) \geq \Delta\omega_{ph}$,

$$P \approx \frac{2}{\sqrt{\pi}} \eta_1 \frac{\gamma'_{tD}}{\max((\gamma'_{tD}+\gamma'_{SE})^2, 1)}. \quad (28)$$

$\eta_1$ is a dimensionless, order of unity coefficient with weak dependence on all frequencies, given by

$$\eta_1 \equiv \frac{\Gamma_{PC}^2 \max(\gamma_{total}^2, \gamma_{tP}^2)\kappa_{DC}^2}{\Delta\omega_{ph}} I_w(\Delta\omega_{ph},\kappa_{DC},\gamma_{total},\gamma_{tP},\Gamma_{PC}). \quad (29)$$

which also indicates how $I_w$ scales with the various frequencies.

**Case 2.** For $\min(\Gamma_{PC},\kappa_{DC}) \geq \Delta\omega_{ph} \geq \max(\gamma_{total},\gamma_{tP})$,

$$P \approx \frac{2}{\sqrt{\pi}} \eta_2 \frac{\gamma'_{tD}}{\max(\gamma'_{tD}+\gamma'_{SE}, 1)} \frac{1}{\Delta\omega'_{ph}}. \quad (30)$$

The dimensionless coefficient $\eta_2$ is also of the order of unity and given by

$$\eta_2 \equiv \Gamma_{PC}^2 \max(\gamma_{total},\gamma_{tP})\kappa_{DC}^2 I_w(\Delta\omega_{ph},\kappa_{DC},\gamma_{total},\gamma_{tP},\Gamma_{PC}). \quad (31)$$

**(Optimal condition for yield)** The above result suggests to optimize $P$ with the following choice of parameters: $\gamma'_{tD} \sim 1 > \gamma'_{SE}$ (i.e., $\gamma_{tD} \sim \gamma_{tP} > \gamma_{SE}$), along with $\min(\Gamma_{PC},\kappa_{DC}) \geq \gamma_{tP} \geq \Delta\omega_{ph}$ (i.e., the condition given in **Case 1**). The condition $\gamma_{tD} \approx \gamma_{tP}$ (or $|A|^2/\Gamma_{PC} \approx |C|^2/\kappa_{DC}$) means the trion emits a photon into the DBR and PC cavity modes with nearly matching rates, at least in order of magnitudes, which imposes a constraint of correlation between the two cavities' parameters. In **Sec. IV**, a numerical evaluation of the integral $I_w$ will be performed for a more detailed study of $P$ as a function of the various parameters.

**(Fidelity)** Next, the fidelity of QST is defined by $F(\alpha,\beta) = \langle\Psi_{ideal}|\hat{\rho}_{output}|\Psi_{ideal}\rangle/P$, where $\Psi_{ideal}$ is given by Eqn. (6), and $\hat{\rho}_{output} = \sum_{\sigma\tau,\sigma'\tau';k_{2D}} \phi_{\sigma\tau}^{output}(k_{2D},\infty)\phi_{\sigma'\tau'}^{output}(k_{2D},\infty)^*$ is the density matrix of the final state. Alternatively, we write

$$F(\alpha,\beta) = \sum_{k_{2D}} F_{k_{2D}} P_{k_{2D}}/P,$$

$$F_{k_{2D}} \equiv \langle\Psi_{ideal}|\sum_{\sigma\tau,\sigma'\tau'}\phi_{\sigma\tau}^{output}(k_{2D},\infty)\phi_{\sigma'\tau'}^{output}(k_{2D},\infty)^*|\Psi_{ideal}\rangle/P_{k_{2D}},$$
(32)

with $F_{k_{2D}}$ the fidelity for a given $k_{2D}$. Using the output amplitudes obtained earlier, one can show that $F_{k_{2D}}$ is a constant independent of $k_{2D}$, and thus obtain

$$F(\alpha,\beta) = F_{k_{2D}}$$
$$= \frac{|\alpha^*A^*(\alpha C + \beta D) + \beta^*A(\alpha D^* + \beta C^*)|^2}{\Big[|(\alpha C + \beta D)|^2 + |(\alpha D^* + \beta C^*)|^2\Big]\Big[|A|^2 + |B|^2\Big]} \quad (33)$$

We note that $F$ given above is a function of the incoming photon state $(\alpha, \beta)$. One could further take the average of $F$ with respect to $(\alpha, \beta)$. Instead, in what follows, we will present the numerical results for both $P$ and $F$, with $F$ being given for representative $(\alpha, \beta)$'s, e.g., $(1, 0)$, $(1/\sqrt{2}, 1/\sqrt{2})$, and etc. The issue of optimization as well as that of minimizing the sensitivity to $(\alpha, \beta)$ will be discussed in **Sec. IV** when we present the numerical result.

### IV. NUMERICAL STUDY

We present numerical results of the yield and fidelity. Effects of damping parameters $\gamma_{SE}$, $\kappa_{DC}$ and $\Gamma_{PC}$, photon bandwidth $\Delta w_{ph}$, and magnitudes of major optical matrix elements $A$ and $C$ will be discussed. For the fidelity $F$, we will examine effects of the ratio $B/A (= D/C)$ between major and minor optical matrix elements, since $F$ depends critically on it.

We start with an estimation of $P$ and $F$ in a typical case. We use $\omega_{ph}$ and the corresponding numerical values of optical matrix elements given in **Sec. II**, namely, $\omega_{ph}$ = $1.6 \cdot 10^5$ GHz, $B/A = D/C \sim 0.04$, $A \sim 45$ GHz, and $C \sim 30$ GHz, with all the matrix elements here taken to be real numbers. Moreover, we take $\gamma_{SE}$ = 1 GHz, $\Delta\omega_{ph}$ = 5 GHz, $\kappa_{DC} = \omega_{ph}/\pi Q$ = 90 GHz corresponding to a cavity $Q \sim 550$, $\Gamma_{PC}$ = 200 GHz corresponding to $Q \sim 250$, and $(\alpha, \beta) = (1, 0)$. The numerical estimation of $P$ and $F$ using the above parameters in Eqns. (27) and (33) gives $P \sim 0.998$ and $F \sim 0.998$.

Next, we discuss the yield $P$ as a function of $\gamma'_{tD}$ (radiative trion damping into DBR cavity modes), $\gamma'_{SE}$ (nonradiative trion damping and radiative trion damping into noncavity modes), and $\Delta\omega'_{ph}$ (photon band width), under the condition given in **Sec. III** for **Cases 1** and **2**, namely, that $\Gamma_{PC}$ and $\kappa_{DC}$ are the largest frequency parameters.

**Figure 5** presents the yield $P$ as a function of the two trion damping rates, $\gamma'_{SE}$ and $\gamma'_{tD}$, for different $\Delta\omega'_{ph}$, with $\Delta\omega'_{ph} = 0.5$ in **Figure 5(a)** and $\Delta\omega'_{ph} = 5$ in **Figure**

**Figure 5** Contour plots of the yield $P$ as a function of trion decay rates $\gamma'_{SE}$ (nonradiative damping and radiative damping into noncavity modes) and $\gamma'_{tD}$ (radiative damping into DBR cavity modes), with different photon band widths: $\Delta\omega'_{ph} = 0.5$ in (a) and $\Delta\omega'_{ph} = 5$ in (b).

**5(b)**. Generally, we see that $P$ decreases with increasing $\gamma'_{SE}$. On the other hand, $P$ varies non-monotonously with $\gamma'_{tD}$, in such a way that in **Figure 5(a)** $P$ attains the maximum $\sim 1$ around $\gamma'_{tD} \sim 1$ for small $\gamma'_{SE}$. In contrast, in **Figure 5(b)**, with $\Delta\omega'_{ph} = 5$, it violates the condition specified for **Case 1** and hence also the optimal condition for $P$. Therefore, $P$ is typically small in this case.

**Figure 6** presents the yield $P$ as a function of $\gamma'_{tD}$ and $\Delta\omega'_{ph}$. The effect of $\gamma'_{SE}$ on yield is studied with $\gamma'_{SE} = 0.1$ in **Figure 6(a)** and $\gamma'_{SE} = 1$ in **Figure 6(b)**. We see that generally $P$ decreases with increasing $\Delta\omega'_{ph}$ but varies non-monotonously with $\gamma'_{tD}$, with the maximum value $\sim 1$ reached at small $\Delta\omega'_{ph}$ in **Figure 6(a)**. In **Figure 6(b)**, because of the relatively large magnitude of $\gamma'_{SE}$, $P$ is overall reduced in comparison to that in **Figure 6(a)**.

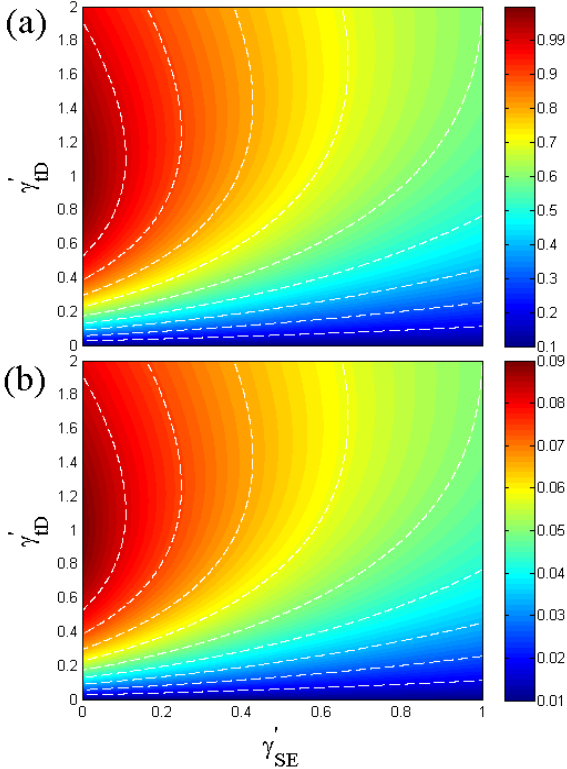

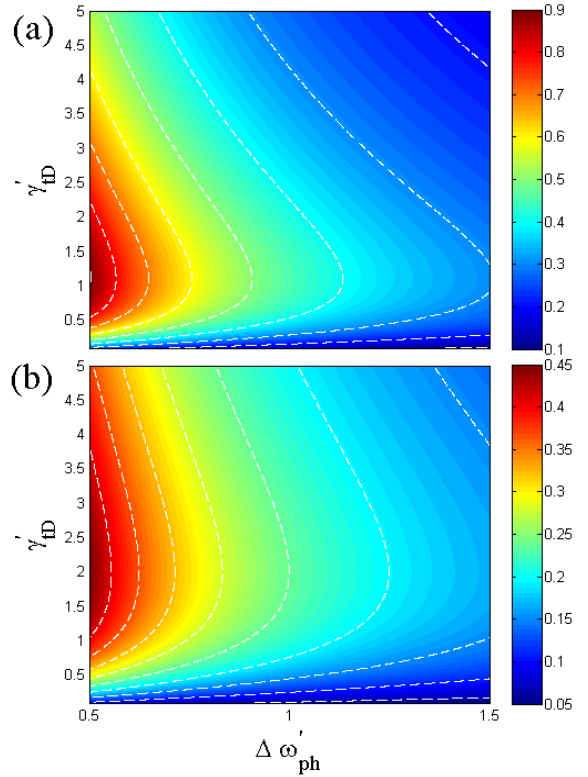

**Figure 6** Contour plots of the yield $P$ as a function of $\Delta\omega'_{ph}$

(photon band width) and $\gamma'_{tD}$ (radiative trion damping into DBR cavity modes). The effect of trion decay rate $\gamma'_{SE}$ (nonradiative damping and radiative damping into noncavity modes) on yield is studied with (a) $\gamma'_{SE} = 0.1$ and (b) $\gamma'_{SE} = 1$.

**Figure 7** presents the yield $P$ as a function of $\gamma'_{SE}$ and $\Delta\omega'_{ph}$, for different $\gamma'_{tD}$'s, with (a) $\gamma'_{tD} = 0.1$, (b) $\gamma'_{tD} = 1$, and (c) $\gamma'_{tD} = 10$. We see that $P$ generally decreases with both increasing $\gamma'_{SE}$ and $\Delta\omega'_{ph}$. Moreover, it varies non-monotonously with $\gamma'_{tD}$, with its value in **Figure 7(a)** and **Figure 7(c)** being overall reduced in comparison to that in **Figure 7(b)** where $\gamma'_{tD} = 1$.

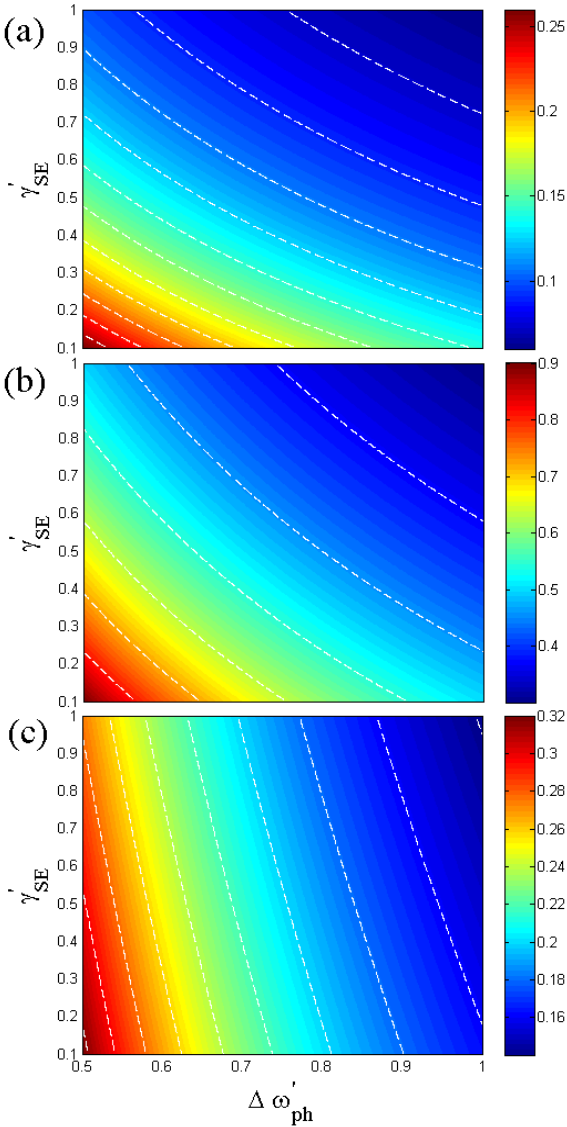

**Figure 7** Contour plots of the yield $P$ as a function of the trion decay rate $\gamma'_{SE}$ (nonradiative damping as well as damping into noncavity modes) and $\Delta\omega_{ph}'$ (photon band width), for different $\gamma'_{tD}$ (trion damping into DBR cavity modes), with (a) $\gamma'_{tD} = 0.1$, (b) $\gamma'_{tD} = 1$, and (c) $\gamma'_{tD} = 10$.

The results shown in **Figures 5-7** indicate that minimizing the trion damping rate $\gamma'_{SE}$ is generally beneficial to the QST yield. In addition, a narrow $\Delta\omega_{ph}'$ that favors the QST to proceed near the resonance condition enhances the yield. Overall, we see that the yield reaches the maximum ~ 1 when $\gamma'_{tD} \sim 1$, $\Delta\omega_{ph}' < 1$ and $\gamma'_{SE} < 1$, confirming the optimal condition given in **Sec. III-2**.

Next, we discuss the fidelity $F$ as a function of $A$, $B$, $C$, $D$, $\alpha$, and $\beta$. In particular, we will examine effects of the ratios $|B/A|$ and $|\beta/\alpha|$, the phase of $D/C$ (denoted as $\phi_{D/C}$), and the phase of $\beta/\alpha$ (denoted as $\phi_{\beta/\alpha}$). We take $A$ and $C$ to be real throughout the discussion.

**Figure 8** presents the fidelity $F$ as a function of $|B/A|$ and $\phi_{D/C}$, with (a) $\alpha = 1, \beta = 0$, (b) $\alpha = 1/\sqrt{2}, \beta = 1/\sqrt{2}$, and (c) $\alpha = 1/\sqrt{2}, \beta = i/\sqrt{2}$. Overall, we see that $F$ decreases with increasing $|B/A|$. On the other hand, while $F$ is independent of $\phi_{D/C}$ in the case of **Figure 8(a)** where the incoming photon signal consists of single circular polarization, in **Figures 8(b)** and **8(c)** $F$ varies periodically in $\phi_{D/C}$, with a relative phase shift by $\pi/2$ between the two figures. These features can be understood in terms of the fidelity formula given in Eqn. (33). By substituting $B = A|B/A|e^{i\phi_{D/C}}$ and $\beta = \alpha|\beta/\alpha|e^{i\phi_{\beta/\alpha}}$ into the formula, we obtain, in the case of $|\alpha| = |\beta| = 1/\sqrt{2}$,

$$F = \frac{|C+|D|\cos\delta|^2}{|C|^2+|D|^2+2C|D|\cos\delta} \frac{|A|^2}{|A|^2+|B|^2} \quad (34)$$

where $\delta = \phi_{D/C} + \phi_{\beta/\alpha}$, which shows that $F$ is indeed periodic in $\phi_{D/C}$, and is shifted in $\phi_{D/C}$ in the presence of a finite $\phi_{\beta/\alpha}$. Moreover, the local maximum and minimum occur at $\delta = n\pi$ and $\delta = (n+1/2)\pi$, respectively, where $n$ = integer. However, although the fidelity varies with $\phi_{\beta/\alpha}$, its overall sensitivity to the incoming signal state can be suppressed by reducing $|B/A|$, as reflected in both Eqn. (34) and **Figure 8**.

Last, **Figure 9** presents the dependence of fidelity $F$ on both $|\beta/\alpha|$ and $\phi_{\beta/\alpha}$, for different combinations of $|B/A|$ and $\phi_{D/C}$, with (a) $|B/A| = 0.04$, $\phi_{D/C} = 0$, (b) $|B/A| = 0.04$, $\phi_{D/C} = \pi/4$, (c) $|B/A| = 0.04$, $\phi_{D/C} = \pi/2$, (d) $|B/A| = 0.4$, $\phi_{D/C} = 0$, (e) $|B/A| = 0.4$, $\phi_{D/C} = \pi/4$, and (f) $|B/A| = 0.4$, $\phi_{D/C} = \pi/2$. We see that $F$ in **Figures 9(a)-9(c)** with $|B/A| = 0.04$ is generally larger than that in **Figures 9(d)-9(f)** with $|B/A| = 0.4$. Moreover, with $|B/A|$ being small in **Figures 9(a)-9(c)**, $F \sim 1$ and is quite robust to the variations in both $|\beta/\alpha|$ and $\phi_{\beta/\alpha}$. In details, $F$ increases with $|\beta/\alpha|$ and reaches the maximum at $|\beta/\alpha| = 1$. Beyond that, although not shown in the graphs, F would be expected to decrease from the maximum when further increasing $|\beta/\alpha|$, since, as Eqn. (33) indicates, $F$ is basically a symmetric function of $\alpha$ and $\beta$. On the other hand, $F$ varies periodically with $\phi_{\beta/\alpha}$ and is shifted by $\phi_{D/C} = \pi/4$ when going from **Figures 9(a)** to **9(b)** or from **9(d)** to **9(e)**, and by $\phi_{D/C} = \pi/2$ when going from **Figures 9(a)** to **9(c)** or from **9(d)** to **9(f)**, with the local maximum and minimum occurring at $\phi_{D/C} + \phi_{\beta/\alpha} = n\pi$ and $\phi_{D/C} + \phi_{\beta/\alpha} = (n+1/2)\pi$, respectively, where $n$ = integer. The periodic behavior displayed here can again be understood in terms of an analysis similar to the earlier one performed for **Figure 8**.

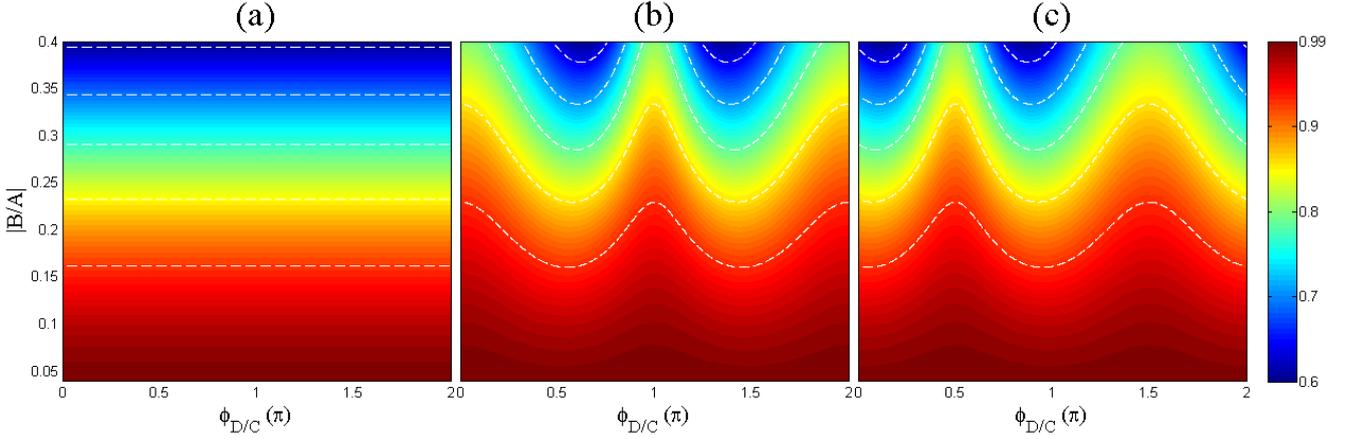

**Figure 8** Contour plots of the fidelity $F$ as a function of $|B/A|$ and the relative phase $\phi_{D/C}$, for different combinations of $\alpha$ and $\beta = 0$: (a) $\alpha = 1, \beta = 0$, (b) $\alpha = 1/\sqrt{2}, \beta = 1/\sqrt{2}$, and (c) $\alpha = 1/\sqrt{2}, \beta = i/\sqrt{2}$.

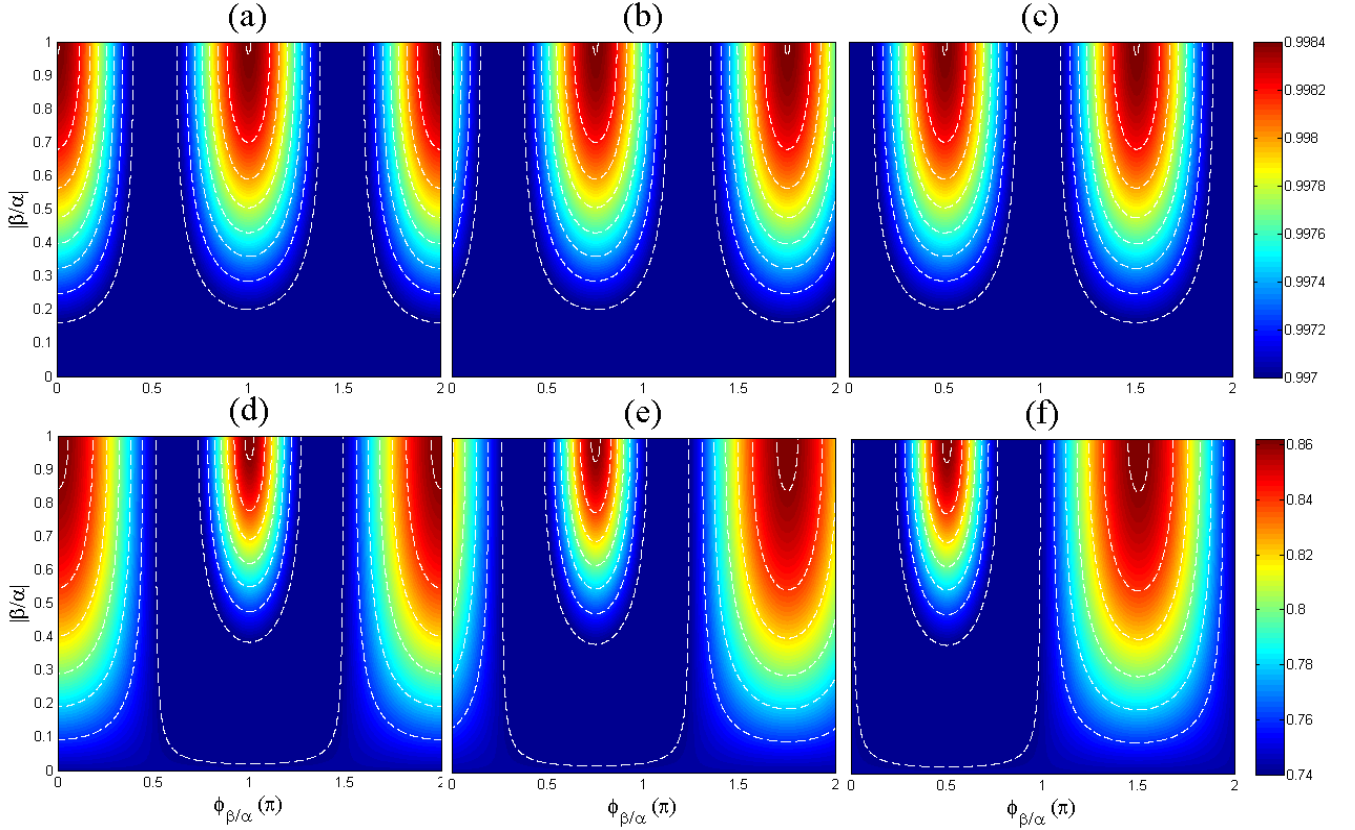

**Figure 9** Contour plot of fidelity as a function of the photon amplitude ratio $|\beta/\alpha|$ and relative phase $\phi_{\beta/\alpha}$. The effect of $|B/A|$ and $\phi_{\beta/\alpha}$ on fidelity is investigated separately, each under different combinations, namely, (a) $|B/A| = 0.04$, $\phi_{D/C} = 0$, (b) $|B/A| = 0.04$, $\phi_{D/C} = \pi/4$, (c) $|B/A| = 0.04$, $\phi_{D/C} = \pi/2$, (d) $|B/A| = 0.4$, $\phi_{D/C} = 0$, (e) $|B/A| = 0.4$, $\phi_{D/C} = \pi/4$, and (f) $|B/A| = 0.4$, $\phi_{D/C} = \pi/2$.

**(Optimal condition for fidelity)** Overall, **Figures 8 and 9** show that the fidelity $F$ depends on $|\beta/\alpha|$ and $\phi_{\beta/\alpha}$ of the incoming signal and $|B/A|$ and $\phi_{D/C}$ of the valley-pair qubit. From the application point of view, they also make the following important suggestion, namely, reduction of the single parameter $|B/A|$ in order to allow for a high fidelity as well as minimization of its sensitivity to $|\beta/\alpha|$, $\phi_{\beta/\alpha}$ and $\phi_{D/C}$. Since, as discussed in **Sec. II-3**, $|B/A|$ is zero for electrons at the band edge and increases with the electron energy, the strategy for favorable fidelity characteristics would therefore be to move electron states to the band edge as close as possible.

### V. CONCLUSION

In summary, we have investigated the valley-photon QST under a hybrid DBR and PC cavity setup that provides both an enhancement of the e-ph interaction and a spatial differentiation between incoming and outgoing photons at the same time. A quantum-mechanical analysis has been performed for the system consisting of electrons, trions, and photons, with the system being open to the environment and allowing for the photons to move in and out. Effects of damping are included. With the analysis, we have derived analytical expressions for the yield and fidelity, which suggest the following condition for an optimized yield: small trion damping rate, narrow photon band width, and nearly matching rates of photon emission by the trion into both cavities, and the following condition for an optimized fidelity: placement of qubit electrons in near-band-edge states. Using realistic qubit and cavity parameters as well as optical matrix elements, a numerical study has also been carried out. A specific example is given with the following parameters: QD size ~ 70nm, photon frequency $\omega_{ph} = 1.6 \cdot 10^5$ GHz corresponding to graphene band gap ~ 0.1 eV, trion damping rate $\gamma_{SE} = 1$

GHz, photon band width $\Delta\omega_{ph}$ = 5 GHz, $Q \sim 550$ for the DBR cavity, and $Q \sim 250$ for the PC cavity. The calculation using the above parameters gives yield and fidelity both near unity.

In conclusion, results of this initial study suggest that the unique valley-polarization correspondence in 2D hexagonal materials such as graphene can be exploited to enable valley-photon QST, with promising characteristics achievable under accessible conditions. Further experimental and theoretical explorations will be important to fully demonstrate such a quantum process as well as realize its full potential for 2D materials-based quantum technologies.

Last, we note that a similar idea of valley-photon QST may be applicable to TMDCs. However, owing to the existence of a strong spin-orbit interaction in TMDCs, spin and valley degrees of freedom are coupled giving rise to a significant distinction between gapped graphene and TMDCs. An extensive work will therefore be required to generalize the valley-photon QST discussed here to TMDCs.

## ACKNOWLEDGEMENTS

We would like to thank MoST, ROC for supporting our work through the Contract No. 103-2119-M-007-007-MY3.

## APPENDIX

Eqn. (17) can be solved by first considering the corresponding homogeneous equation, an eigenvalue problem for the matrix

$$H_0 = \begin{pmatrix} a & \sqrt{|A|^2+|B|^2} \\ \sqrt{|A|^2+|B|^2} & b \end{pmatrix}, \quad (A-1)$$

$$a \equiv \omega_{trion} - i\gamma_{total}, \; b \equiv \omega_{PC} - i\Gamma_{PC}.$$

The solutions are given by the following eigenvalues

$$\lambda_1 = \frac{(a+b)+\sqrt{(a-b)^2+4(|A|^2+|B|^2)}}{2},$$

$$\lambda_2 = \frac{(a+b)-\sqrt{(a-b)^2+4(|A|^2+|B|^2)}}{2}, \quad (A-2)$$

and eigenvectors

$$|\varphi_1\rangle = \begin{pmatrix} \phi_{11} \\ \phi_{12} \end{pmatrix} = \begin{pmatrix} -\dfrac{-a+b+\sqrt{(a-b)^2+4(|A|^2+|B|^2)}}{2\sqrt{|A|^2+|B|^2}} \\ 1 \end{pmatrix},$$

$$|\varphi_2\rangle = \begin{pmatrix} \phi_{21} \\ \phi_{22} \end{pmatrix} = \begin{pmatrix} -\dfrac{-a+b-\sqrt{(a-b)^2+4(|A|^2+|B|^2)}}{2\sqrt{|A|^2+|B|^2}} \\ 1 \end{pmatrix}. \quad (A-3)$$

Next, we include the inhomogeneous part $f(t)$. We express $f(t)$ in terms of the eigenvectors of $H_0$:

$$f(t) = c_1(t)\begin{pmatrix} \phi_{11} \\ \phi_{12} \end{pmatrix} + c_2(t)\begin{pmatrix} \phi_{21} \\ \phi_{22} \end{pmatrix}, \quad (A-4)$$

with the expansion coefficients given by

$$\begin{cases} c_1(t) = \dfrac{\phi_{22}}{\phi_{11}\phi_{22}-\phi_{12}\phi_{21}}\left[C\phi_{\sigma_+K}^{DC}(t)+D\phi_{\sigma_-K}^{DC}(t)\right] \\ c_2(t) = \dfrac{-\phi_{12}}{\phi_{11}\phi_{22}-\phi_{12}\phi_{21}}\left[C\phi_{\sigma_+K}^{DC}(t)+D\phi_{\sigma_-K}^{DC}(t)\right] \end{cases} \quad (A-5)$$

It can be verified that the solution to Eqn. (17) is given by

$$\begin{pmatrix} \phi_K^{tri}(t) \\ \phi_{K1}^{PC}(t) \end{pmatrix} = -i\Bigg[ \frac{\phi_{22}}{\phi_{11}\phi_{22}-\phi_{12}\phi_{21}}\begin{pmatrix} \phi_{11} \\ \phi_{12} \end{pmatrix}\int_0^t dt' e^{-i\lambda_1(t-t')}C\phi_{\sigma_+K}^{DC}(t')$$

$$-\frac{\phi_{12}}{\phi_{11}\phi_{22}-\phi_{12}\phi_{21}}\begin{pmatrix} \phi_{21} \\ \phi_{22} \end{pmatrix}\int_0^t dt' e^{-i\lambda_2(t-t')}C\phi_{\sigma_+K}^{DC}(t')$$

$$+\frac{\phi_{22}}{\phi_{11}\phi_{22}-\phi_{12}\phi_{21}}\begin{pmatrix} \phi_{11} \\ \phi_{12} \end{pmatrix}\int_0^t dt' e^{-i\lambda_1(t-t')}D\phi_{\sigma_-K}^{DC}(t')$$

$$-\frac{\phi_{12}}{\phi_{11}\phi_{22}-\phi_{12}\phi_{21}}\begin{pmatrix} \phi_{21} \\ \phi_{22} \end{pmatrix}\int_0^t dt' e^{-i\lambda_2(t-t')}D\phi_{\sigma_-K}^{DC}(t')\Bigg]$$

(A-6)

Using Eqn. (14) for the DBR cavity photon amplitude, it gives

$$\phi_{K1}^{PC}(t)$$

$$= -i\sqrt{c\kappa_{DC}}\phi_0^{ph}\frac{\phi_{12}\phi_{22}}{\phi_{11}\phi_{22}-\phi_{12}\phi_{21}}\left[\alpha C+\beta D\right]$$

$$\left[\int_0^t dt' e^{-i\lambda_1(t-t')} - \int_0^t dt' e^{-i\lambda_2(t-t')}\right]$$

$$\int_{-\infty}^{\infty}\frac{d\omega}{2\pi}e^{\frac{-i(\omega-\omega_{ph})x_0}{c}}e^{\frac{-(\omega-\omega_{ph})^2}{2\Delta\omega_{ph}^2}}\frac{e^{i\omega t'}-e^{-i(\omega_{DC}-i\kappa_{DC}-iC)t'}}{(\omega+\omega_{DC}-i\kappa_{total})}$$

(A-7)

Now, substituting the above result into Eqn. (24) and evaluating the resultant integral, we arrive at the final state amplitude when the QST is completed:

$$\lim_{x_0 \to -\infty} \phi_{K1}^{output}\left(k_{2D}, t \to \infty\right)$$

$$= -e^{-i\omega_{output}(k_{2D})t} T_k \sqrt{c\kappa_{DC}} \phi_0^{ph} \frac{\phi_{12}\phi_{22}}{\phi_{11}\phi_{22} - \phi_{12}\phi_{21}} [\alpha C + \beta D]$$

$$\frac{(\lambda_1 - \lambda_2)}{\left(\omega_{output}(k_{2D}) - \lambda_1\right)\left(\omega_{output}(k_{2D}) - \lambda_2\right)\left(\omega_{DC} - i\kappa_{total} - \omega_{output}(k_{2D})\right)}.$$

(A-8)

This leads to the result in Eqn. (25).

# References

[1] A. Rycerz, J. Tworzydlo, and C.W. J. Beenakker, Nat. Phys. **3**, 172 (2007).
[2] D. Xiao, W. Yao, and Q. Niu, Phys. Rev. Lett. **99**, 236809 (2007).
[3] R. V. Gorbachev, J. C. W. Song, G. L. Yu, A. V. Kretinin, F. Withers, Y. Cao, A. Mishchenko, I. V. Grigorieva, K. S. Novoselov, L. S. Levitov, A. K. Geim, Science **346**, 448 (2014).
[4] H. Zeng, J. Dai, W. Yao, and X. Cui, Nat. Nanotechnol. **7**, 490 (2012).
[5] K. F. Mak, K. He, J. Shan, and T. F. Heinz, Nat. Nanotechnol. **7**, 494 (2012).
[6] Q. H. Wang, K. Kalantar-Zadeh, A. Kis, J. N. Coleman and M. S. Strano, Nat. Nanotechnol. **7**, 699 (2012).
[7] Y. J. Zhang, T. Oka, R. Suzuki, J. T. Ye and Y. Iwasa, Science **344**, 725 (2014).
[8] W. Yao, D. Xiao, and Q. Niu, Phys. Rev. B **77**, 235406 (2008).
[9] K. S. Novoselov, A. K. Geim, S. V. Morozov, D. Jiang, Y. Zhang, S. V. Dubonos, I. V. Grigorieva, and A.A. Firsov, Science **306**, 666 (2004); A. K. Geim and K. S. Novoselov, Nature Mater. **6**, 183 (2007).
[10] Y. Zhang, Y.-W. Tan, H. L. Stormer, and P. Kim, Nature **438**, 201 (2005).
[11] A. H. Castro Neto, F. Guinea, N. M. R. Peres, K. S. Novoselov, and A. K. Geim, Rev. Mod. Phys. **81**, 109 (2009).
[12] K. F. Mak, C. Lee, J. Hone, J. Shan, and T. F. Heinz, Phys. Rev. Lett. 105, 136805 (2010).
[13] G. Y. Wu and N. -Y. Lue and L. Chang, Phys. Rev. B **84**, 195463 (2011).
[14] N. Rohling and G. Burkard, N. J. Phys. **14**, 083008 (2012); N. Rohling, M. Russ and G. Burkard, Phys. Rev. Lett. **113**, 176801 (2014).
[15] G. Y. Wu and N. -Y. Lue and Y. -C. Chen, Phys. Rev. B **88**, 125422 (2013).
[16] A. Kormányos, V. Zólyomi, N. D. Drummond, and G. Burkard, Phys. Rev. X **4**, 011034 (2014).
[17] Y. Wu, Q. Tong, G.-B. Liu, H. Yu, and W. Yao, Phys. Rev. B **93**, 045313 (2016).
[18] M. T. Allen, J. Martin, and A. Yacoby, Nature Commun. **3**, 934 (2012).
[19] X.-X. Song, D. Liu, V. Mosallanejad, J. You, T.-Y. Han, D.-T. Chen, H.-O. Li, G. Cao, M. Xiao, G.-C. Guo, and G.-P. Guo, Nanoscale, **7**, 16867 (2015)
[20] K. Wang, T. Taniguchi, K. Watanabe, and P. Kim, arXiv:1610.02929 (2016)
[21] C. G. Yale, F. J. Heremans, B. B. Zhou, A. Auer, G. Burkard, and D. D. Awschalom, *Nature Photonics* **10**, 184 (2016); and references therein.
[22] F. Xia, H. Wang, D. Xiao, M. Dubey and A. Ramasubramaniam, Nature Photon. **8**, 899 (2014).
[23] X. Liu and V. M. Menon, IEEE J. Quantum Electron. **51**, (2015).
[24] X. Liu, T. Galfsky, Z. Sun, F. Xia, E. C. Lin, Y. H. Lee, S. Kena-Cohen and V. M. Menon, Nature Photon. **9**, 30 (2015).
[25] X. Gan, Y. Gao, K. F. Mak, X. Yao, R. -J. Shiue, A. van der Zande, M. E. Trusheim, F. Hatami, T. F. Heinz, J. Hone, and D. Englund, Appl. Phys. Lett. **103**, 181119 (2013).
[26] G. Y. Wu and N. -Y. Lue, Phys. Rev. B **86**, 045456 (2012).
[27] Y. Rikitake, H. Imamura and H. Kosaka, J. Phys. Soc. Jpn. 76, 114004 (2007).
[28] D. A. Lidar, I. L. Chuang, and K. B. Whaley, Phys. Rev. Lett. **81**, 2594 (1998); M. Mohseni, and D. A. Lidar, *ibid.* 94, 040507 (2005).
[29] J. I. Cirac, P. Zoller, H. J. Kimble, and H. Mabuchi, Phys. Rev. Lett. **78**, 3221 (1997); T H. J. Briegel, W. Dur, J. J. Cirac, and P. Zoller, Phys. Rev. Lett. **81**, 5932 (1998); L.-M. Duan, M. D. Lukin, J. J. Cirac, and P. Zoller, Nature **414**, 413 (2001).
[30] H. J. Kimble, Nature 453, 1023 (2008). [quantum repeater]
[31] B. B. Blinov, D. L. Moehring, L. -M. Duan and C. Monroe, Nature 428, 153 (2004).
[32] S. D. Jenkins, S. -Y. Lan, T. A. B. Kennedy and A. Kuzmich, Nature **438**, 833 (2005).
[33] A. Stute, B. Brandstatter, K. Friebe, T.E. Northup and R. Blatt, Nature Photon. **7**, 219 (2013).
[34] C. Kurz, M. Schug, P. Eich, J. Huwer, P. Műller and J. Eschner, Nature Commun. **5**, 5527 (2014).
[35] M. A. Sillanpää, J. I. Park and R. W. Simmonds, Nature **449**, 438 (2007); M. Hua, M. -J. Tao and F. -G. Deng, Scientific Reports 6, 22037 (2016); P. Xu, X. -C. Yang, F. Mei and Z. -Y. Xue, Scientific Reports 6, 18695 (2016).
[36] E. Togan, Y. Chu, A. S. Trifonov, L. Jiang, J. Maze, L. Childress, M. V. G. Dutt, A. S. Sørensen, P. R. Hemmer, A. S. Zibrov, and M. D. Lukin, Nature **466**, 7307 (2010).
[37] N. Y. Yao, L. Jiang, A. V. Gorshkov, Z, -X. Gong, A. Zhai, L. -M. Duan and M. D. Lukin, Phys. Rev. Lett. **106**, 040505 (2011).
[38] H. Kosaka, H. Shigyou, Y. Mitsumori, Y. Rikitake, H. Imamura, T. Kutsuwa, K. Arai and K. Edamatsu, Phys. Rev. Lett. **100**, 096602 (2008).
[39] H. Kosaka, T. Inagaki, Y. Rikitake, H. Imamura, Y, Mitsumori and K, Edamatsu, Nature **457**, 702 (2009).
[40] M. A. Nielsen and I. L. Chuang, Quantum comupation and Quantum Information (Cambridge University Press, Cambridge, 2003), and references therein.
[41] A. Kitaev, Ann. Phys., **303**, 2 (2003); C. Nayak, S. H. Simon, A. Stern, M. Freedman, S. D. Sarma, Rev. Mod. Phys. **80**, 1083 (2008).
[42] H. Ryu, S. Kim, H. Park, J. Hwang, Y. Lee and J. Kim Appl. Phys. Lett. **80**, 3883 (2002).
[43] Y. Akahane, T. Asano, B. Song and S. Noda, Nature **425**, 944 (2003).
[44] Y. Wakayama, A. Tandaechanurat, S. Iwamoto and Y. Arakawa, Opt. Express **16**, 21320 (2008).
[45] R. Nascimento, J. da Rocha Martins, R. J. C. Batista, and H. Chacham, J. Phys. Chem. C **119**, 5055 (2015).
[46] H. -C. Cheng, N. -Y. Lue, Y. -C. Chen, and G. Y. Wu, Phys. Rev. B **89**, 235426 (2014)
[47] O. Painter and Srinivasan, Phys. Rev. B 68, 035110 (2003).